\documentclass{aa}  
\usepackage[colorlinks]{hyperref}
\usepackage{natbib}
\usepackage{scrhack} 
\hypersetup{%
    colorlinks=true,
    linkcolor=[RGB]{0,0,255}, 
    linkbordercolor={0 0 1},
    filecolor=[RGB]{0,0,255},      
    urlcolor=[RGB]{0,0,255},
    citecolor=[RGB]{0,0,255},
}
\usepackage{float}
\usepackage{physics}
\usepackage{amsmath}
\usepackage{eucal}
\usepackage{amsfonts}
\usepackage{amssymb}
\usepackage{graphicx}
\usepackage{txfonts}
\usepackage{longtable}  
\usepackage{lscape}
\usepackage{xcolor}
\newcommand{\REV}[1]{{\color{black}#1}}
\newcommand{\newREV}[1]{{\color{black}#1}}

\definecolor{seagreen}{rgb}{0.190, 0.525, 0.361}

\begin{document}

\title{Binary stars in the Milky Way nuclear stellar cluster}


\author{Arn Marklund,
      \inst{1}\fnmsep\inst{2}
      Ross P.~Church \inst{2}
      \and Alessandro A. Trani \inst{3,4,5}
      }

\institute{Observatoire Astronomique de Strasbourg, Université de Strasbourg, CNRS UMR 7550, 11 rue de l’Université, F-67000 Strasbourg, France\\ \email{arn.marklund@astro.unistra.fr} \and
        Division of Astrophysics, Department of Physics, Lund University, Box 118, SE-221 00 Lund, Sweden
         \and 
         Niels Bohr International Academy, Niels Bohr Institute, Blegdamsvej 17, 2100 Copenhagen, Denmark
          \and
          National Institute for Nuclear Physics – INFN, Sezione di Trieste, I-34127, Trieste, Italy
          \and
          Departamento de Astronom\'ia, Facultad Ciencias F\'isicas y Matem\'aticas, Universidad de Concepci\'on, Avenida Esteban Iturra, Casilla 160-C, Concepci\'on, 4030000, Chile
         }

\date{Received XXX; accepted YYY}

\abstract
{\REV{Intermediate mass galaxies, including the Milky Way, typically} host both a supermassive black hole (SMBH) and a nuclear stellar cluster (NSC). Binary systems residing in an NSC evolve dynamically via frequent close encounters with surrounding stars and secular processes related to the SMBH.}
{\REV{Although} the evolution of very soft and very hard binaries can be predicted semi-analytically, the situation is more \REV{complex} for binaries that lie near the hard-soft boundary. We aim to follow their evolution throughout the age of the NSC ($\sim 10~\mathrm{Gyr}$), determine their evolutionary outcomes and the effects on a binary population in the NSC.}
{By employing numerical simulations of three-body encounters between binary systems and a tertiary star, while also considering the secular evolution in the form of von Zeipel-Lidov-Kozai oscillations and tidal dissipation, we follow the evolution of moderately soft and hard binaries ($0.03-2.5~\mathrm{au}$) of initial masses $\lesssim 2~\mathrm{M}_\odot$ at galactocentric radii of 0.1 and 0.3\,pc.}
{We find that inward migration caused by three-body encounters leads to the destruction of binaries through mergers and evaporation, while outward migration is a pathway to retaining intact binaries for $\gtrsim 10~\mathrm{Gyr}$. All binaries that remain intact are hard and circular, but the outcomes for binaries initially at the hard-soft boundary are highly stochastic. From the destroyed binaries, we find that \textit{i)} $\sim0.3\%$ of evaporated binaries fall into the SMBH's loss cone \textit{ii)} $\gtrsim1\%$ of mergers occur late enough to be observed as blue straggler stars (BSSs) on the main-sequence or as recently evolved red giants \textit{iii)} $\sim1\%$ of the mergers originating at $0.1~\mathrm{pc}$ merge at orbits completely confined to the inner arcsec of the NSC and \textit{iv)} $\lesssim 80\%$ of collisions between a field star in the NSC and one of the binary stars, leads to a subsequent merger with the other binary star; a three-body pile up (3BPU). These 3BPUs are relatively common within the first 1-2 Gyr but stagnate after that, and could serve as a way to form more massive BSSs.}
{We predict that a small, but possibly substantial, fraction of binaries in the NSC originate closer to the SMBH than their present-day orbits. Conversely, we expect evaporated binary stars and merger products in the form of BSSs close to the SMBH that originated further out in the NSC. The binaries' and merger products' orbits confined to the inner arcsec of the NSC \newREV{end up there after $\gtrsim 300\,\mathrm{Myr}$ and have circular orbits. They are therefore unlikely to be related to the formation of the S-stars or G-objects, but suggest that the inner arcsec is contaminated with BSSs from earlier star formation events.}  
}

\keywords{Stellar dynamics -- Binaries -- Stellar collisions -- }

\maketitle

\section{Introduction} \label{Introduction}

Galaxies with a stellar mass $\lesssim 10^{10}\mathrm{M}_\odot$\footnote{Galaxies exceeding this stellar mass only host a SMBH.} typically host a nuclear stellar cluster (NSC) in their most central region \citep[][and references therein]{Neumayer+2020}. At the center of these structures there is typically a supermassive black hole \citep[SMBH, ][]{NeumayerWalcher2012, Nguyen+2019} with a mass between $10^{6-9}\mathrm{M}_\odot$ \citep[see e.g.][]{Kormendy&Ho2013}. The Galactic Center (GC) of the Milky Way is no exception; the radio source Sagittarius A* (Sgr A*) is associated with a SMBH of $4\times 10^6~\mathrm{M}_\odot$, which in turn is surrounded by a $2.5\times 10^7~\mathrm{M}_\odot$ NSC \citep[see e.g.][]{Schodel+2003, Schodel+2014,Eisenhauer+2005, Ghez+2005, Ghez+2008, Gillessen+2009, Gillessen+2017, Alexander&Pfuhl2014,Boehle+2016,GravCollab2018}.

Very close to Sgr A* there exists a young population of stars, usually referred to as the S-cluster, which indicates that recent star formation has taken place \citep[e.g.][]{Eisenhauer+2005,Genzel+2010,Schodel+2020}. \REV{So far, finding binaries among these stars have not proven very successful \citep{Chu_2023}. Only a small number of main-sequence (MS) binaries have been detected in the Galactic Center\footnote{There have, however, been some observations of X-ray binaries; see, for example, \citet{Muno+2005a, Muno+2005b, Hailey+2018}.}, and those identified are all massive \citep{Martins+2006, Pfuhl+2014}. In contrast, little is known about a less massive and fainter MS population in the NSC}. Recently, \citet{Pei_ker_2024} presented the detection of a lower-mass spectroscopic binary ($2.80\pm0.50~\mathrm{M}_\odot$ and $0.73\pm0.14~\mathrm{M}_\odot$ components) within the S-cluster. One idea, which is further motivated by this detection, is that these lower-mass binaries could be the progenitors of the G-objects; dust-enshrouded self-gravitating cloud-like stellar sources \citep{Ghez+2005,Gillessen2012G2,Ciurlo+2020}. All binary systems in the \newREV{inner parts of the NSC form a hierarchical triple system with the SMBH.} As such, the SMBH acts as a distant perturber that induces eccentricity- and inclination oscillations in the binaries: the so called von~Zeipel-Lidov-Kozai \citep[ZLK,][]{Zeipel, Lidov, Kozai} mechanism. These oscillations can be extreme enough that they drive the binaries to near parabolic orbits \newREV{where they may merge} because of the finite size of the constituent stars. This process is one of the leading candidate mechanisms for the formation of the G-objects (\citealp{Prodan+2015, 2016Stephen&Naoz, StephanNaoz2019, Ciurlo+2020}, see also \citealp{g2_2012ApJ...750...58B,g2_2012ApJ...755..155S,g2_2013ApJ...776...13B,g2_2014ApJ...783...31S,g2_2014ApJ...789L..33D,g2_2015ApJ...806..197M,g2_2016ApJ...819L..28B,trani2016b,g2_2016MNRAS.455.4388C,g2_2017A&A...602A.121Z,g2_2018MNRAS.478.3494C,g2_2018MNRAS.479.5288B, trani2019a}). Another mechanism responsible for producing binary mergers is frequent encounters between binaries and surrounding stars \citep[see][and this study]{Hamers&Samsing2019, Samsing+2019, Michaely&Perets2019, Young&Hamers2020}.

While \citet{Prodan+2015, 2016Stephen&Naoz, StephanNaoz2019} all look at the dynamical evolution of binaries via the ZLK mechanism, their simulations are confined to the inner $0.1~\mathrm{pc}$ of the \newREV{NSC} \REV{because} further out, the general relativity precession (GRP) timescale becomes short enough that ZLK may be suppressed \citep{Ford+2020,Liu+2015,Naoz+2013a, Naoz2016}. \newREV{Additionally, far out in the NSC, other} relevant dynamical timescales become long enough that they are not too different from the Galactic field \citep[see e.g.][]{Rose}. This leaves a small region ($\sim 0.1-0.3~\mathrm{pc}$) where the dynamical evolution of a binary is dominated by its interactions with nearby stars and where associated dynamical processes occur on a timescale shorter than or comparable to the age of the NSC ($\sim 10~\mathrm{Gyr}$). 

\REV{In this paper, we aim to i) highlight the different outcome drivers that determine the fate of binaries and ii) relate different types of binary remnants to the variety of objects that exist in the NSC, with extra focus on binary mergers.} \REV{The rest of the paper is structured as follows: In Section \ref{Dynamics in the NSC} we describe our model for the NSC (\ref{sect: NSC model}) and the dynamical processes considered for this study (\ref{Intro: 3body encounters}-\ref{Tides}). In Section \ref{Method} we describe the binary population we consider (\ref{sect: Stellar Populations}) and how we evolve our binaries (\ref{sect: Evolving a binary} - see also Appendix \ref{Accept/Reject}). Section \ref{Results} shows the results we obtain which are divided into mergers (\ref{sect:Mergers}), evaporation (\ref{sect:Evaporation}) and the remaining outcomes (\ref{sect:Hard binaries})\footnote{We include a short discussion on tidal disruption events in Appendix \ref{Appendix TDE}}. This is followed by a general discussion in Section \ref{Discussion}, including a more focused discussion on binary mergers (\ref{Discussion: Binary mergers}), and comparisons with other relevant studies (\ref{Discussion: Other studies}). Finally, our conclusions are presented in Section \ref{Conclusion}.
}
\section{\REV{Dynamics in the nuclear stellar cluster} }\label{Dynamics in the NSC}

The orbits of binary systems in the NSC are modified by various processes. Because all binaries form a hierarchical triple system with the SMBH at the center of the NSC, both general relativity (GR) and ZLK oscillations affect their evolution \citep{Prodan+2015, Naoz2016, 2016Stephen&Naoz, StephanNaoz2019}. Moreover, as the stellar density in the NSC itself is very high, encounters with surrounding stars cannot be ignored and drive both evaporation \citep{Heggie1975,Rose} and mergers \citep{Hamers&Samsing2019}. In the following sections, we outline our approach to dealing with these mechanisms.

\subsection{NSC model} \label{sect: NSC model}

The presence of a SMBH has a number of effects on the surrounding stars and the NSC. As the cluster ages and relaxes, it reaches a steady-state characterised by a cuspy power-law stellar density profile (see eqn. \ref{Eq: NSC density}) with an exponent $\alpha = 1.75$, typically referred to as a Bahcall-Wolf profile \citep{Bahcall&Wolf1976}\footnote{The Bahcall-Wolf profile naturally emerges in single mass systems.}
\begin{equation}\label{Eq: NSC density}\rho\left(r_{\bullet}\right)=\rho_0\left(\frac{r_{\bullet}}{r_0}\right)^{-\alpha}.
\end{equation}
\REV{$r_\bullet$ is the distance to the SMBH, and we adopt the scale density} $\rho_0=1.35\times10^6~\mathrm{M}_\odot\mathrm{pc}^{-3}$ and $r_0=0.25\mathrm{pc}$ following \citet{Genzel+2010}. The gravitational field in the inner parsec of the NSC is dominated by the SMBH, therefore, stars inside this radius tend to move on Keplerian orbits. Additionally, the resulting one dimensional velocity dispersion $\sigma$ related to the density distribution via the Jeans equation, scales with the mass of the SMBH $M_\bullet$

\begin{equation} \label{Eq: Velocity Dispersion}
    \sigma (r_\bullet) = \sqrt{\frac{GM_\bullet}{r_\bullet(1+\alpha)}}.
\end{equation}
\REV{where $G$ is the gravitational constant}
We adopt the Bahcall-Wolf profile of $\alpha = 1.75$ in these simulations, but note that observations of stars in the NSC suggest slightly shallower profiles with values between $\alpha \sim 1.25-1.5$ \citep{Gallego-Cano+2018}.
\REV{
\subsection{Dynamical three-body encounters} \label{Intro: 3body encounters}

As a binary interacts with a third star (tertiary), there is an exchange of energy and angular momentum that changes the separation and eccentricity of the binary. The separation is inversely proportional to the binary's internal binding energy, and as such, if the binary loses (gains) energy, its separation grows (shrinks). Typically, one refers to the growing and shrinking as softening and hardening, respectively, which is related to the binary's hardness ratio; the ratio of its binding energy to the average kinetic energy of the surrounding stars:
\begin{equation} \label{Eq: hardness ratio}
    h = \frac{2E_{\rm bin}}{\bar{m_3} \sigma^2}
\end{equation}
where $\Bar{m_3}$ and $\sigma$ are the average mass and velocity dispersion of the surrounding stars, respectively. A binary is considered hard if $h>1$, and soft if $h<1$. Based on the seminal work of \citet{Heggie1975}, it has been shown that there is a tendency for soft binaries to soften and for hard binaries to harden with time; a trend usually referred to as Heggie's law. A soft binary is therefore expected to soften until it eventually evaporates. In the inner regions of the NSC, the velocity dispersion is on the order of a few hundreds $\rm km\,s^{-1}$, see eqn. (\ref{Eq: Velocity Dispersion}), which is more than an order of magnitude higher than for a typical globular cluster \citep[see e.g.][]{Ivanova+2005}. For a typical population of binaries, e.g. those we find in the field \citep{Raghavan2010, MoeDiStefano2017}, the majority would be considered soft in the NSC and would therefore be expected to evaporate \citep[see e.g.][]{Prodan+2015, 2016Stephen&Naoz, StephanNaoz2019, Panamarev+2019}.}

\subsection{von Zeipel-Lidov-Kozai mechanism} \label{ZLK mechanism}

As mentioned previously, all binaries in the NSC form hierarchical triple systems with the SMBH. The SMBH exerts small but consistent gravitational perturbations on the binary's orbit, which can induce periodic oscillations between the binary's eccentricity and inclination. It is possible to describe the long-term evolution of such a system with Hamiltonian perturbation theory. We base our implementation on the quadrupole approximation and the equations derived in \citet{Naoz+2013a}, and refer the reader to \citet{Naoz2016} for a review. More details on our implementation can also be found in Appendix \ref{Appendix: Secular Processes}.

The ZLK mechanism operates under resonance; any perturbations of this resonance can work to suppress the ZLK oscillations. GRP is one such mechanism that is highly relevant in the NSC. If the timescale for GRP is shorter than the quadrupole timescale, ZLK is suppressed \citep[e.g.][]{Naoz+2013a, StephanNaoz2019, Ford+2020}. The quadrupole timescale is given in eqn. (\ref{eqn: tquad}), where $m_{bin}$ is the binary mass, $M_\bullet$ is the mass of the SMBH, $P_\mathrm{inner}$ ($P_\bullet$) is the orbital period for the inner (outer) orbit and $e_\bullet$ is the orbital eccentricity of the outer orbit.

\begin{equation} \label{eqn: tquad}
    t_\mathrm{quad} \approx \frac{16}{30 \pi} \frac{m_{bin} + M_\bullet}{M_\bullet} \frac{P_\bullet^2}{P_\mathrm{inner}}(1-e_\bullet^2)^{3/2}
\end{equation}

The ratio between this timescale and the GRP timescale can be estimated from \citet[][eqn. 60]{Naoz2016} and the dimensionless parameter $\epsilon_\mathrm{GR}$ introduced in \citet{Liu+2015}. These two expressions are given in eqn. (\ref{t_GR vs t_quad}) and eqn. (\ref{eps_GR}) respectively.
\begin{equation} \label{t_GR vs t_quad}
\frac{t_\mathrm{GRP}}{t_\mathrm{quad}} = \epsilon_\mathrm{GR}^{-1}\left(1-e_\mathrm{bin}^2 \right)
\end{equation}

\begin{equation} \label{eps_GR}
\epsilon_\mathrm{GR} = \frac{3Gm_\mathrm{bin}^2 a_\bullet^3\left(1-e_\bullet^2\right)^{3/2}}{a_\mathrm{bin}^4 c^2 M_\bullet}
\end{equation}
The ratio $t_\mathrm{GRP} / t_\mathrm{quad}$ depends on the separation and eccentricity of the inner orbit $a_\mathrm{bin}$ and $e_\mathrm{bin}$, respectively, the binary mass $m_\mathrm{bin}$, the separation and eccentricity of the outer orbit $a_\bullet$ and $e_\bullet$, respectively, and the mass of the SMBH $M_\bullet$. \REV{$G$ is the gravitational constant and $c$ is the speed of light}. ZLK oscillations are only possible if this ratio is greater than one, and as long as the binary-SMBH system is hierarchical: $a_\mathrm{inner}/a_\mathrm{outer} < 0.1$.

\subsection{Tidal dissipation} \label{Tides}

The statistical tendency for binaries to thermalise as they interact with surrounding stars leads to more highly eccentric orbits \citep{Heggie1975, Heggie1975_Clusters}. Similarly, the ZLK mechanism also provides a pathway for producing highly eccentric orbits. When stars in an eccentric binary are at periapsis, tidal forces coupling to the internal structure of the stars (e.g. convective motions) leads to orbital angular momentum transfer and energy dissipation to the stars. As a consequence, the binary orbit circularises and the binary becomes harder; this may lead to the binaries either merging or remaining intact for a $\sim$ Hubble time, as they become virtually stable against evaporation. We base the following on \citet{Mardling&Aarseth2001} and refer the reader to this paper for more detailed explanations regarding tidal dissipation. Importantly, two distinct dissipation mechanisms are considered, as referred to by the original authors: normal (or equilibrium) and chaotic dissipation.

\subsubsection{Equilibrium tidal evolution} \label{Section: Equil Tides}

\REV{Equilibrium tides arise from a quasi-hydrostatic deformation of the star \citep{Zahn1977}. 
This causes a net torque, in which the orbital angular momentum and orbital energy is dissipated into the stars, leading to more circular orbits with smaller separations \citep{Triaud+2017}}. In the equilibrium model, the eccentricity changes as

\begin{multline} \label{Eq: dedt Circ}
    \frac{de}{dt} = -\frac{2}{4\pi^2}\left(\frac{C_1 Q_1^2}{w_1^4} q(1+q) + \frac{C_2 Q_2^2}{w_2^4} \frac{1}{q} \left(1 + \frac{1}{q}\right)\left( \frac{r_2}{r_1}\right)^8\right) \\ 
    \times \frac{f_e(e)}{(p_0(1+e_0))^8},
\end{multline}
where
\begin{equation} \label{Eq: f(e)}
    f(e) = \frac{9\pi}{5} e(1-e^2)^{3/2}\left(1 + \frac{15}{4}e^2 + \frac{15}{8}e^4 + \frac{5}{64}e^6\right),
\end{equation}
$C_i = w_i/2\pi \mathcal{Q}_i$, $\mathcal{Q} = 10^4$ is some nominal value used for the damping timescales defined in \citet{Mardling&Aarseth2001}, $w_i$ is the oscillation frequency, and $Q_i$ are the associated overlap integrals \citep{Press&Teukolsky1977}. These are all structure constants dependent on the polytrope of choice, which we take to be $n=3$ as a reasonable model for the internal structures of low-mass MS stars. The values are tabulated in Table 1 of \citet{Mardling&Aarseth2001}; for the $l=2$ $f-$modes, we have $w_1 = w_2 = 8.175$ in units of $Gm_*/r_*^3$ and $Q_1 = Q_2 = 0.2372$. $q = m_2/m_1$ is the mass fraction where $m_1$ is the dominant star; since all our stars are of the same polytrope, the dominant star that absorbs most of the tidal energy is always the one with the largest radius, $r_1$, \citep{Mardling&Aarseth2001}. We assume conservation of orbital angular momentum $p = p_0 \frac{1+e_0}{1+e}$ to obtain the separation $a = \frac{p r_1}{1-e}$.

\subsubsection{Chaotic tidal evolution} \label{Chaotic Region}

\REV{Low-mass MS stars and giants have convective envelopes. At very close periapsis separations, a coupling between the equilibrium tide and the internal convective motions of the star(s) may result in damping via a turbulent viscosity \citep{Mardling1995a,Mardling1995b,Mardling&Aarseth2001}. The theory presented in \citep{Mardling1995a,Mardling1995b} suggests that this tidal evolution becomes chaotic; the exchange in energy between the orbit and the stars is unpredictable and allows the tides to build up to a significant fraction of the binding energy of the star.}

Binaries that are very eccentric with a periapsis radius of only a few stellar radii lie in what \citet{Mardling&Aarseth2001} refer to as the chaotic region (see their Section 2.3). As the binary loses orbital energy and circularises, once enough energy has been lost, the binary leaves the chaotic region and proceeds to circularise normally \citep[see Section \ref{Section: Equil Tides} and e.g. Figures 3 \& 8 in][]{Mardling&Aarseth2001}.

For the purposes of this study, we are mostly interested in how much the binary circularises in the chaotic region. We assume that the chaotic phase (typically $\gtrsim100\,{\rm yr}$) is short compared to the time between encounters (on average $\gtrsim 10^4\,{\rm yr}$ for the binaries that undergo chaotic tides in our simulations) and hence that a binary that has entered the chaotic tidal regime can be treated as exiting instantaneously. Following \citet{Mardling&Aarseth2001} we obtain the final eccentricity $e$ from\footnote{A factor of $(1+e)^5$ is missing from this equation in the original paper.}

\begin{equation} \label{Eq: Chaotic Eccentricity}
    \frac{(1+e)^{2.4}}{1-e}(1+e)^5 = \left( \frac{p_0(1+e_0)}{C}\right)^5 \left( \frac{\omega^2}{1+q}\right)^2
\end{equation}

where the initial conditions are the eccentricity $e_0$ and the periapsis separation scaled by the stellar radius $r_1$, $p_0 = a_0(1-e_0)/r_1$. Here $q$ is the mass ratio of the secondary to the primary, $C=1.836$, and $\omega$ is the oscillation frequency structure constant for a polytrope of $n=3$, values can be found in \citet[][table 1]{Mardling&Aarseth2001}. The solution to eqn. (\ref{Eq: Chaotic Eccentricity}) is obtained numerically, and we get the semi-major axis directly from the initial state by assuming constant orbital angular momentum.

\section{Method} \label{Method}

In this section, we outline the initial conditions, our binary population, and our routine for combining the different processes described in Section \ref{Dynamics in the NSC}.

\subsection{Initial stellar and binary population} \label{sect: Stellar Populations}
The stars in the NSC are predominantly old, with roughly $80\%$ being formed more than 5 Gyr ago \citep{Pfuhl+2011, Schodel+2020, NSC_history2023}. Furthermore, \citet{Pfuhl+2011} found that the old population of giant branch stars mainly consists of stars corresponding to zero-age MS masses of $1-1.2~\mathrm{M}_\odot$, which would put solar mass stars right at the MS turn-off. 
We are interested primarily in binaries that may be contributing to the old visible population today, which from expectations would be binaries with $m_1 \lesssim 1\mathrm{M_\odot}$ \citep{Pfuhl+2011, Schodel+2020} and located near the hard/soft boundary \citep[H/SB, i.e. $h\sim1$, see eqn. \ref{Eq: hardness ratio}. See, also e.g.][]{Panamarev+2019,StephanNaoz2019, Rose}. Additionally, the evolution of binaries near the H/SB is the most uncertain (consider, for example, Heggie's law), which makes the employment of direct N-body methods very useful if not necessary. Consistent with the turn-off mass, we take the age of the NSC to be 10 Gyr and the primary to be between $0.9-1.0~\mathrm{M}_\odot$.

\begin{figure}[H]
    \centering
    \includegraphics[width=\linewidth]{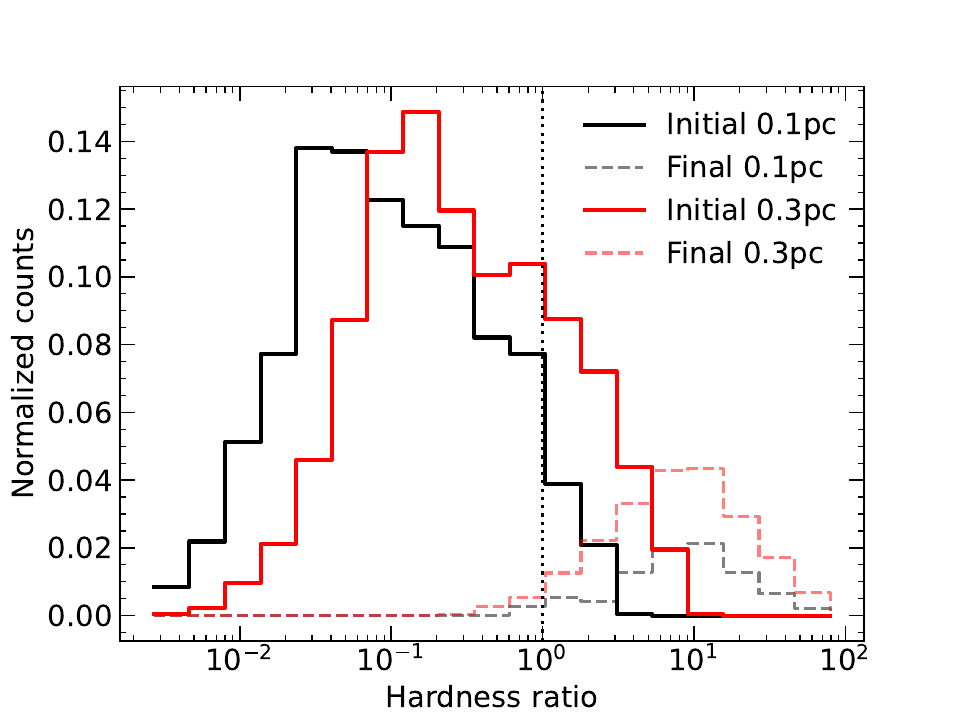}
    \caption{\newREV{Initial (final) distribution of hardness ratio for binaries originating at $0.1$ and $0.3~\mathrm{pc}$ in solid (dashed) black and red lines, respectively. The vertical dotted line marks the H/SB. Each bin is normalised to the total sum.}}
    \label{fig:h-distr}
\end{figure}

The mass ratio, period (thereby also separation), and eccentricity for the binaries are all sampled from the distributions in \citet{MoeDiStefano2017} using the publicly available package \texttt{COSMIC} \citep{Breivik+2020}. We consider secondary masses between $0.1-1.0~\mathrm{M}_\odot$. We know very little about the binary population in the NSC and especially how a primordial population might have looked like. A binary population in the NSC with properties similar to the binaries found in the solar neighbourhood \citep[see][]{Raghavan2010, MoeDiStefano2017} would, for the majority, be considered soft in the NSC due to the high velocity dispersion (see eqn. \ref{Eq: hardness ratio}) and should inevitably evaporate \citep{Prodan+2015,2016Stephen&Naoz,StephanNaoz2019, Panamarev+2019}. \REV{More generally, binary period distributions in dense stellar systems are poorly constrained even for present day distributions \citep[see, e.g.][]{Ji_2015,Muller-Horn+2025}. The present day population in the NSC should have a narrower distribution than in globular clusters because of the higher number density and velocity dispersion, but not much can be said regarding the primordial population. Because of the lack of observational constraints and theoretical guidance, we assume the primordial binary population to resemble the field population of short to relatively long period binaries.} We put the upper limit on the binary periods at 1000 days and let the lower limit remain at the default value of 1.4 days (for a $2~\mathrm{M}_\odot$ binary this roughly corresponds to $2.5~\mathrm{au}$ and $0.03~\mathrm{au}$, respectively); harder binaries are going to be extremely rare and will likely merge within a relatively short time, and softer binaries should evaporate rather quickly \citep[see e.g.][]{2016Stephen&Naoz}. \newREV{Figure \ref{fig:h-distr} shows the initial distribution of hardness ratios at $0.1$ and $0.3\,\mathrm{pc}$ in black and red solid lines, respectively. The vertical dotted line marks the H/SB.}

The stellar and binary properties ensure that we can avoid complex stellar evolution. Instead, we set up a mass-radius interpolation scheme using stellar tracks produced with \texttt{STARS} \citep{Stars0,STARS} of masses between $0.1-3.0~\mathrm{M}_\odot$ to also accommodate potential mass gain in collisions. In cases of collisions, we assume mass and momentum conservation and no mixing; the remnant's MS lifetime is determined by interpolating the helium mass fraction at the time of merger for the binary star to the corresponding age for the remnant. The helium fraction interpolation scheme is also produced using masses and ages from stellar tracks produced with \texttt{STARS}.

\subsection{Evolving a binary} \label{sect: Evolving a binary}

In this section, we outline our general treatment for the full dynamical evolution of our binaries. 

\subsubsection{Stellar dynamical interactions} 
\label{Stellar Interactions}

We work under the assumption that the primordial binary fraction is sufficiently small that we can neglect binary-binary and higher-order encounters. In other words, we are only considering three-body encounters, in which the following outcomes are possible: 1) collisions, 2) scatterings, 3) exchanges, and 4) captures, where we form transient triples. It is in turn possible for the binary to be destroyed following one of these encounters, in which it can 5) ionise/evaporate or 6) merge. Additionally, because we only consider the mass of the SMBH and by extension Keplerian orbits, and no stellar evolution beyond the MS, we are forced to terminate the dynamical evolution of the binary if it 7) unbinds from the SMBH (ejected), 8) crosses the SMBH Roche limit, and 9) the primary evolves. We outline the conditions for each of the possible outcomes 1-9 to be fulfilled below. Subscripts 1 and 2 refer to the primary and secondary of the initial binary, respectively, and subscript 3 denotes the tertiary.

\begin{enumerate}
    \item A collision occurs when $r_p / (r_i + r_3) < 1$, where $r_p$ is the periapsis and $r_i$, $r_3$ are the radii of the two colliding stars ($i=1,2$), respectively. All collisions are assumed to conserve mass and momentum. While destructive collisions do occur in the GC, they are mostly confined to the inner $0.01~\mathrm{pc}$ \citep{Rose+2023}. 

    \item A scatter (also referred to as a flyby in the literature) is defined as resulting in a bound binary ($E_{1,2} < 0$) while the tertiary is both unbound ($E_{12,3} > 0$) and moving away ($\mathbf{x_3} \cdot \mathbf{v_3} > 0$) from the binary centre of mass. If these two constraints are fulfilled and the tertiary is further away than $10a$, $a$ being the instantaneous separation, we consider the scatter to be concluded.

    \item An exchange is similar to a scatter, but instead of the tertiary being unbound, one of the binary constituents ($i$) becomes unbound from the common centre of mass of the remaining two stars $E_{i,j3} > 0$ where $i = 1,2$ and $j = 3-i$. As before, if $\mathbf{x_i} \cdot \mathbf{v_i} > 0$ and the distance between star $i$ and the common centre of mass of the other two stars is greater than $10a$, the exchange is fulfilled.

    \item When a tertiary is captured by the binary we form a transient triple, which require all three stars to be bound to their common centre of mass, e.g. $E_{i,COM} < 0$ for $i = 1,2,3$. If we produce a transient triple, and the tertiary orbit is longer than $10^3$ days we assume the encounter to be a scatter; the tertiary will spend most of its time at apoapsis where it very easily should be unbounded. If the orbit is shorter than $10^3$ days we resolve the encounter by an additional 10 tertiary orbital periods, if at after this point nothing has happened to the binary we also assume a scatter (however, we find no such instances).

    \item In an ionisation event all stars must be unbound, e.g. $E_{i,j} > 0$ for $i = 1,2,3$ and $j = 1,2,3 \neq i$. Furthermore, they also have to be moving away from the tertiary centre of mass $\mathbf{x_i} \cdot \mathbf{v_i} > 0$ for $i = 1,2,3$. In order to reduce computational requirements we also consider binaries that have $a > 100\,\mathrm{au}$ to inevitably evaporate (ionise) which is therefore also considered as a stopping condition. 

    \item A merger can be caused by the hardening of an encounter (e.g. scatter or collision), due to ZLK oscillations or due to Roche Lobe overflow. In the two former cases, the same stopping condition as in a collision is used, while for the Roche lobe overflow we instead require $R_{1,2} > f(q)a(1-e)$, where $f(q)$ denotes Eggleton's formula \citep{Eggleton1983} for the Roche lobe and is given by\footnote{Just as before, the fraction is given by the secondary over the primary, $q = m_2/m_1$.}
    \begin{equation*}
    f(q) = \frac{0.49q^{2/3}}{0.6q^{2/3} + \ln{1+q^{1/3}}}.
    \end{equation*}

    \item Ejection implies that the binary itself unbinds from the SMBH and simply requires $E_{COM}^{\mathrm{bin-SMBH}} > 0$. 

    \item If the binary ventures inside the SMBH Roche limit it evaporates, with a possibility of a tidal disruption event (TDE). We require $\frac{a_2}{a_1} < \left(\frac{3M_\bullet}{M_{1,2}}\right)^{1/3}\frac{1+e_1}{1-e_2}$.

    \item If the binary becomes older than its primary's expected MS lifetime, we terminate its evolution to avoid computing complex stellar evolution. The star's MS lifetimes are interpolated from stellar tracks produced with \texttt{STARS}.
\end{enumerate}

All termination constraints naturally work as criteria to run three-body interactions, ZLK, and tidal evolution.

\begin{figure*}[htbp!]
    \centering 
    \includegraphics[width = \linewidth, trim=4cm 7cm 2cm 6.5cm, clip]{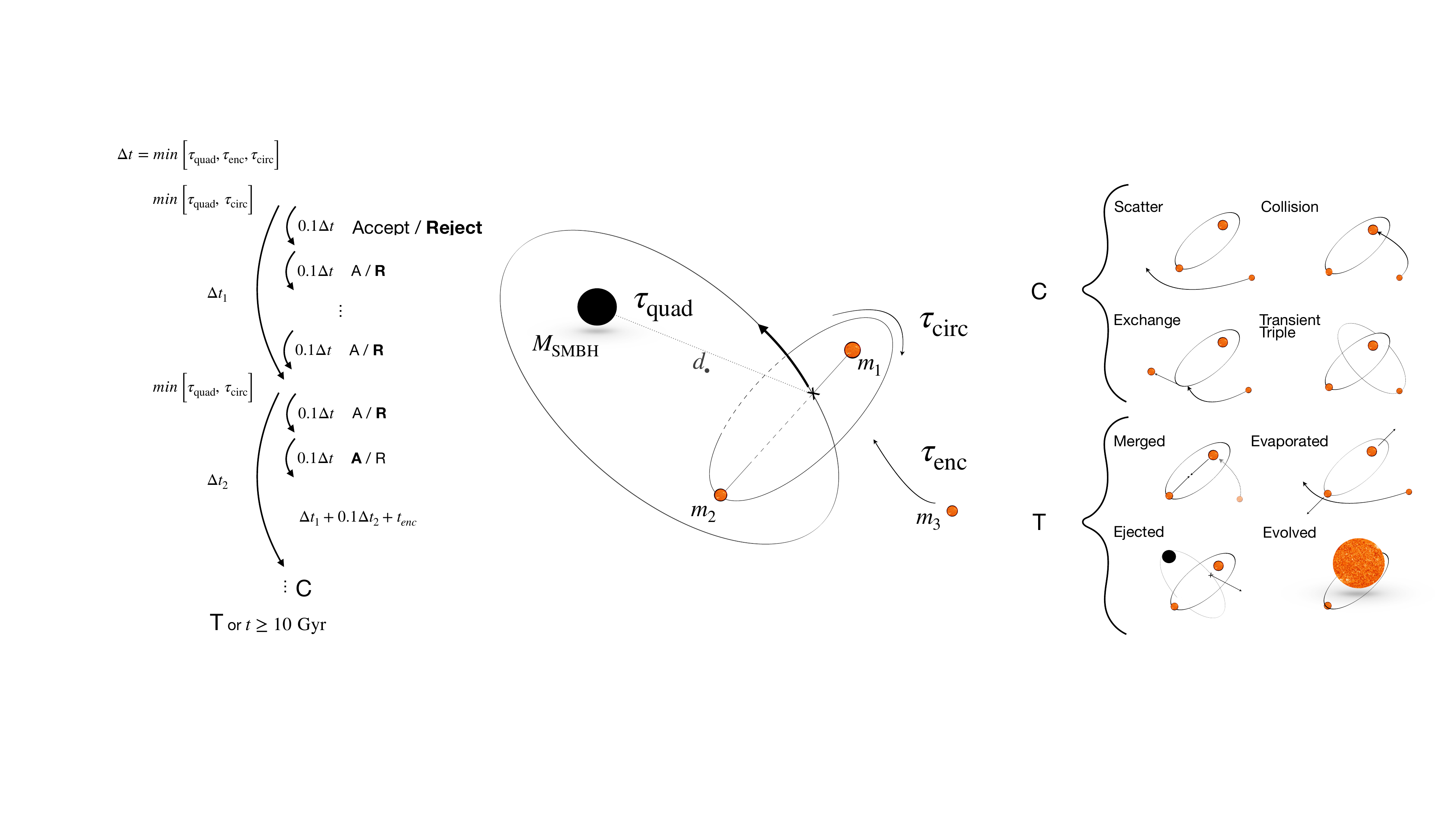}
    \caption{\newREV{Flow chart of our routine which is described in Section \ref{sect: Algorithm}. The central sketch shows the overall setup of the three-body encounters; a binary ($m_1,~ m_2$) on an orbit at a distance of $d_\bullet$ around the SMBH ($M_\mathrm{SMBH}$) interacting with a tertiary ($m_3$). To the right of this sketch we show the possible outcomes from a three-body interaction. Group C contains interactions in which the binary remains intact and bound to the SMBH, and as a consequence will have a Continued evolution. Conversely, the evolution is Terminated by outcomes in group T. To the left of the sketch we illustrate how a binary is evolved. A timestep $\Delta t$ is set by the shortest timescale of ZLK (quadrupole approximation), encounters and circularisation (tides). The secular process with the shortest timescale $\min[\tau_\mathrm{quad},\tau_\mathrm{circ}]$ is then run for $\Delta t$. At every step $0.1\Delta t$, a three body encounter is sampled with the accept reject method described in Appendix \ref{Accept/Reject}; if accepted the three-body encounter is carried out with \texttt{tsunami}, if rejected we proceed to the next $0.1\Delta t$ and sample again. After a time $\Delta t$ has passed, or after a three-body encounter, a new $\Delta t$ is calculated. In this example, the encounter occurs after $\Delta t_1 + 0.1\Delta t_2 + t_\mathrm{enc}$. This process is then continued until an outcome in group T occurs, or until the simulation time exceeds $10\,\mathrm{Gyr}$.}}
    \label{fig:flow chart}
\end{figure*}

\subsubsection{Three-body encounters} \label{three-body encounters}

The numerical integration of three-body encounters is performed using \texttt{tsunami} \citep{tsunami}. In total, we simulate more than $6\times10^7$ encounters, with roughly 1 in every 150 000 encounters leading to something different than a scatter. The SMBH is not included in the three-body encounters, and as such it is only the interactions between the binary and the tertiary we simulate. Furthermore, we do not follow the ZLK evolution in \texttt{tsunami}, but rather integrate the quadrupole equations of motion described in Section \ref{ZLK mechanism}. Finally, the implementation of the equilibrium tides described in Section \ref{Tides} \citep[see also][]{Mardling&Aarseth2001}, is consistent with that implemented in \texttt{tsunami}. As such, even though the duration of any meaningful (equilibrium) tidal evolution typically far exceeds the time a binary spends in a \texttt{tsunami} encounter, we are not "missing" any contributions to its tidal evolution. The order in which these three mechanisms (three-body encounters, ZLK and tides) are considered is explained in Section \ref{sect: Algorithm}.

For simplicity, we neglect the influence of the SMBH's tidal field on the dynamics of three-body encounters in this work. The SMBH can impact encounter durations, merger rates, and the final orbital properties of surviving binaries, as demonstrated in studies on three-body scatterings of stellar-mass black holes in the presence of a SMBH \citep{trani2019b,trani2020IAUS,trani2024a}. A comprehensive investigation of its effects on stellar encounters is left for future studies. \REV{Nevertheless, the binary's orbit around the SMBH is allowed to change: we update the outer orbit with the velocity kicks received in any three-body encounter. Hereafter, we refer to this process as migration.}

\begin{figure*}[htbp!]
    \centering
    \includegraphics[width =0.93\textwidth, trim=0cm 1cm 0cm 2cm, clip]{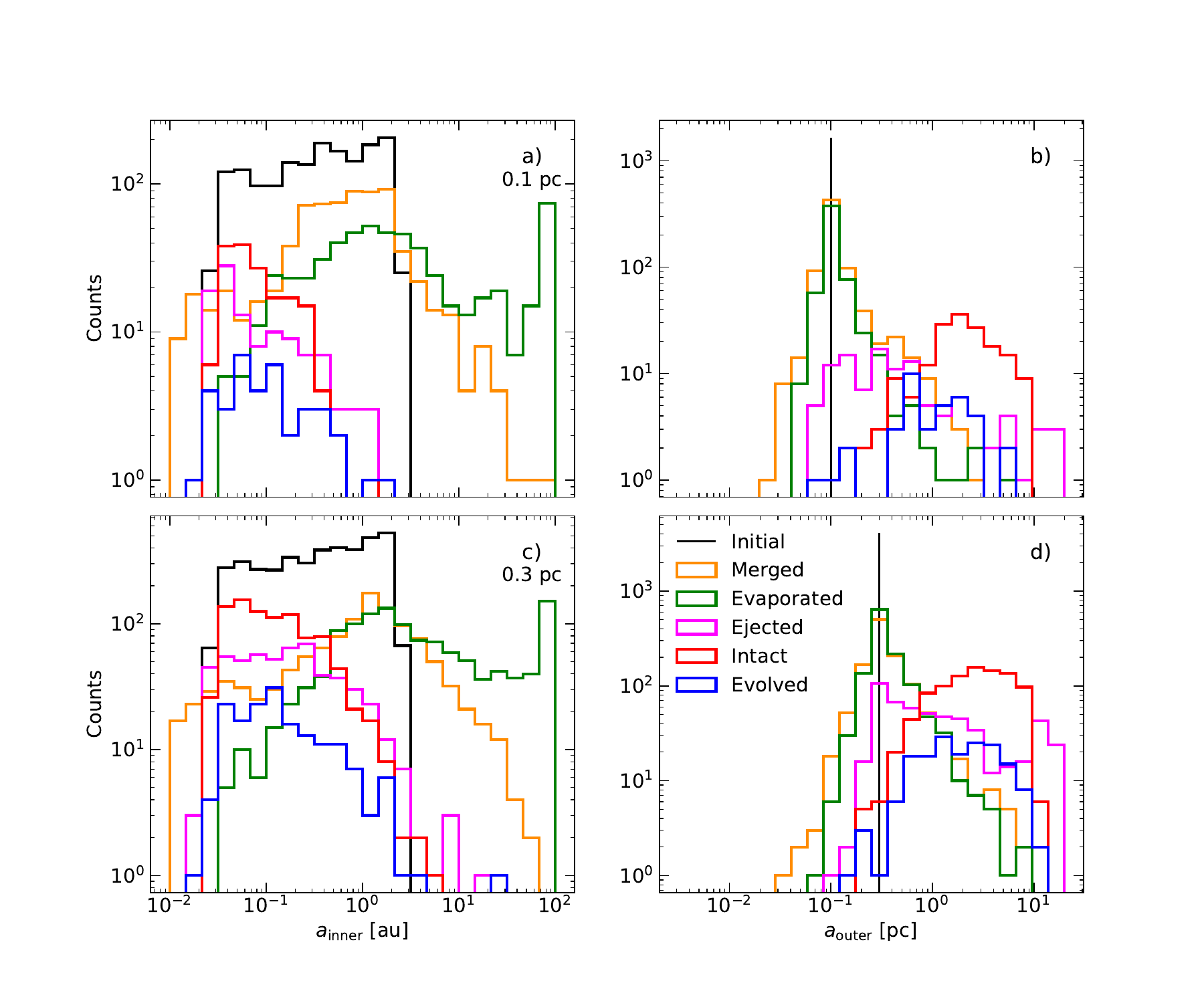}
    \caption{{\it Left:} Inner semi-major axes distribution of the initial population (black) and the corresponding final outcomes (colours). Yellow lines refer to binary mergers; green lines to binaries that evaporate; magenta lines to binaries that are ejected (unbound from the SMBH); red lines to intact binaries (i.e. binaries that survive to 10 Gyr); and blue lines to evolved binaries (stellar evolution of the primary occurs before binary destruction and by 10 Gyr). In the cases of evaporation and mergers, the properties shown in the Figure refers to the binary just before either destruction mechanism takes place. Ejection corresponds to the region the binaries were ejected from and not where they end up. The top row corresponds to binaries starting with $a_\bullet = 0.1~\mathrm{pc}$ from the SMBH, the bottom row to $a_\bullet = 0.3~\mathrm{pc}$.  {\it Right:} Outer semi-major axis distributions. The vertical black lines represent the initial populations, the boxes the final outcomes coloured in the same way.  
    }
    \label{fig: sma distributions}
\end{figure*}

\subsubsection{Routine} \label{sect: Algorithm}

We introduce a hierarchical set-up that is governed by the timescales of the different processes. The ZLK quadrupole timescale $t_\mathrm{quad}$ is given by eqn. (\ref{eqn: tquad}), the instantaneous timescale for encounters $t_\mathrm{enc}$, is given in Appendix \ref{Accept/Reject} by taking the inverse of eqn. (\ref{Eq: Encounter Rate}), and for a tidal dissipation timescale $t_\mathrm{circ}$ we refer the reader to eqns. 20--25 in \citet{Mardling&Aarseth2001}.

The basis of our routine lies in the three-body encounters; only these are resolved with \texttt{tsunami}, and the secular evolution (ZLK and tides) is computed in-between encounters. The hierarchy of the set-up looks like this:

\begin{itemize}
\item {\it Chaotic tidal evolution}: If a binary lies within the chaotic region, its tidal evolution is almost always fast enough that tidal processes dominate \citep{Mardling&Aarseth2001}. We neglect the time taken for these dynamical tides to damp out and follow the procedure laid out in section \ref{Chaotic Region} to obtain the post-chaos parameters.

\item {\it Secular evolution}: Next we compare the timescales for the ZLK mechanism ($t_{\rm quad}$; see section \ref{ZLK mechanism}) and the equilibrium tide ($t_{\rm circ}$; see section \ref{Tides}). The secular process with the shortest timescale is selected.

\item {\it Close encounters}: We obtain the instantaneous encounter timescale $t_{\rm enc}$ that encompasses all dynamically significant encounters (see Appendix \ref{Accept/Reject}).
\end{itemize}
\vspace{-0.1cm}
\REV{\noindent We first sample the binaries with \texttt{COSMIC} and place them on an orbit around the SMBH with orbital elements drawn according to the distributions in table \ref{tab:ICs}. The impact parameter, velocity and mass for the tertiary are drawn as described in Appendix \ref{Setting up a tertiary}. We calculate a time-step $\delta t = \mathrm{min}(t_\mathrm{quad},~t_\mathrm{circ},~t_\mathrm{enc})$ and compute the secular process with the shortest timescale for a duration of $\delta t$ given that\footnote{If none of the secular processes are possible, we only consider three-body encounters.}:
\begin{enumerate}
    \item[\textit{i}] If circularisation: $t_\mathrm{circ} <1~\mathrm{Gyr}$ and $e_\mathrm{inner} > 0.002$. If $t_\mathrm{circ} >1~\mathrm{Gyr}$ and the eccentricity $\leq 0.002$, the binary is declared circularised \citep[similar to][]{Mardling&Aarseth2001}. 
    \item[\textit{ii}] If ZLK: The mutual inclination must be large enough for the torques to arise ($\gtrsim 40~\deg$), and GRP cannot dominate (e.g. eqn. \ref{t_GR vs t_quad} must be smaller than unity). 
\end{enumerate}
\vspace{-0.1cm}
At fixed intervals of $0.1\Delta t$ we sample a three-body encounter with the Accept/Reject method described in Appendix \ref{Accept/Reject}: if rejected, the secular evolution continues and we take another step $0.1\delta t$; if accepted, we resolve the interaction with \texttt{tsunami} and determine the outcome based on the conditions defined in Section \ref{three-body encounters}. For outcomes 1-4, we update the binary's inner and outer orbits, and the process is then restarted. For outcomes 5-9, the dynamical evolution of the binary is terminated.} \newREV{A flow chart of the routine is included in Figure \ref{fig:flow chart}, which also shows a sketch of the setup and the different outcomes.}

\begin{table}[H]
    \centering
    \caption{\REV{Distributions for the orbital elements not drawn from \texttt{COSMIC}. If the inner or outer orbit is not specified, the distribution is used for both orbits (but sampled independently)}.}
    \begin{tabular}{clc}
        Symbol & Definition & Distribution \\ \hline
        $a_\mathrm{outer}$ & Outer separation & $0.1, 0.3~\mathrm{pc}$ \\
        $e_\mathrm{outer}$ & Outer eccentricity & thermal \\
        $i$ & inclination & uniform in $\cos i$ \\
        $\omega$ & Initial argument of periastron & uniform \\
        $\Omega$ & Longitude of ascending node & uniform \\
        $M$ & Mean anomaly  & uniform \\
        
    \end{tabular}
    \label{tab:ICs}
\end{table}


\section{Results}\label{Results}

Figure \ref{fig: sma distributions} shows an overview of the outcomes from our simulations, and we base the following discussion on these results. The upper (bottom) row refers to the binaries originally set at $0.1~\mathrm{pc}$ ($0.3~\mathrm{pc}$) of the SMBH. The left-hand side column shows the inner semi-major axis while the right-hand side column shows the outer semi-major axis. The black lines correspond to the initial conditions; for the outer orbits these are vertical lines with a height equivalent to the total number of binaries considered at respective distance (1654 at $0.1~\mathrm{pc}$ and 4082 at $0.3~\mathrm{pc}$). Also, see Table \ref{tab: final outcomes} for the total number number of binaries for each outcome. In the subsections below, we go through Figure \ref{fig: sma distributions} line by line, starting with mergers and evaporation (orange and green lines, respectively), followed by the ejected (magenta), intact (red) and evolved (blue) populations.

\begin{figure}
    \centering    
    \includegraphics[width=0.95\linewidth, trim=0cm 0cm 0cm 1cm, clip]{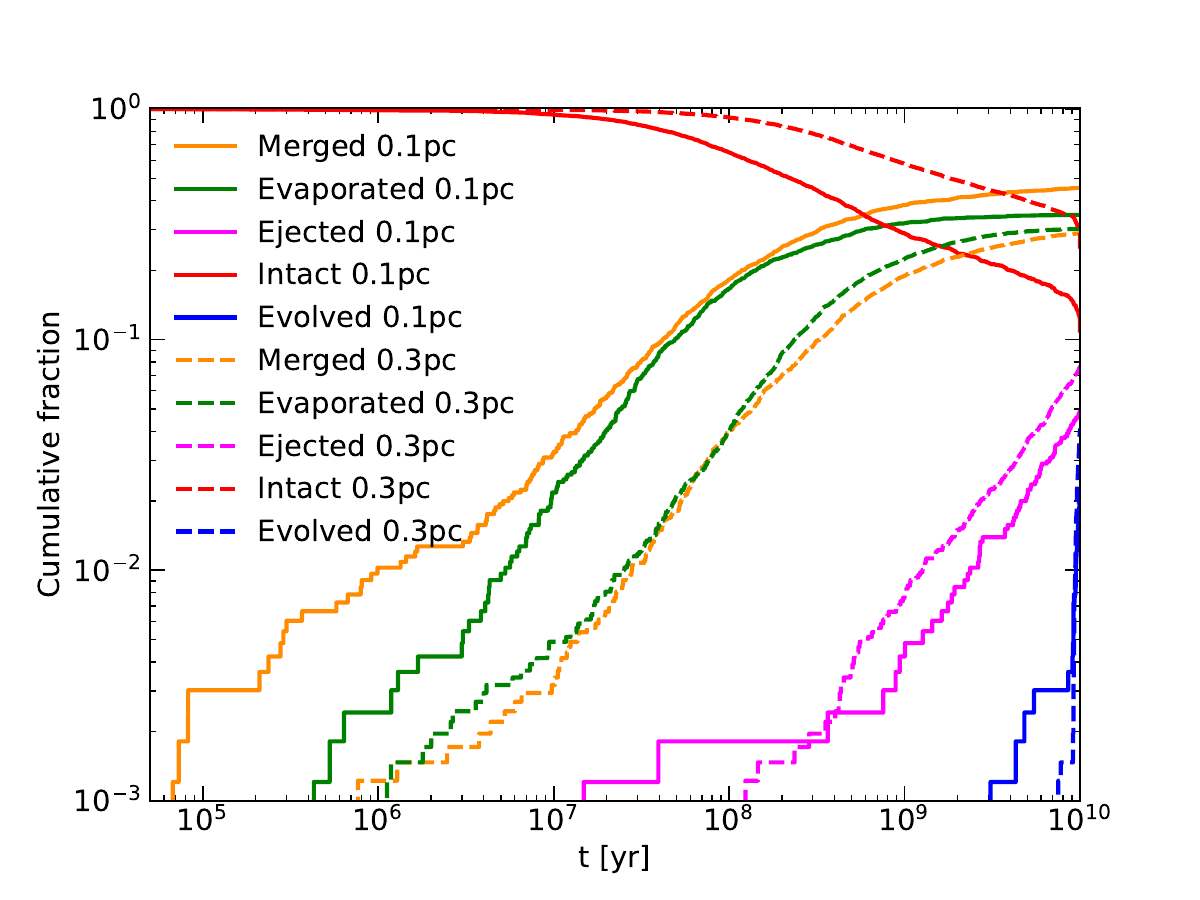}
    \caption{\REV{Cumulative distribution of outcomes as a function of time. The outcomes are colour-coded the exact same way as in Figure \ref{fig: sma distributions}. The solid (dashed) lines refer to binaries initially $0.1~\mathrm{pc}$ ($0.3~\mathrm{pc}$) away from the SMBH.}}
    \label{fig: evolution1}
\end{figure}

\begin{figure}
    \centering
    \includegraphics[width=\linewidth]{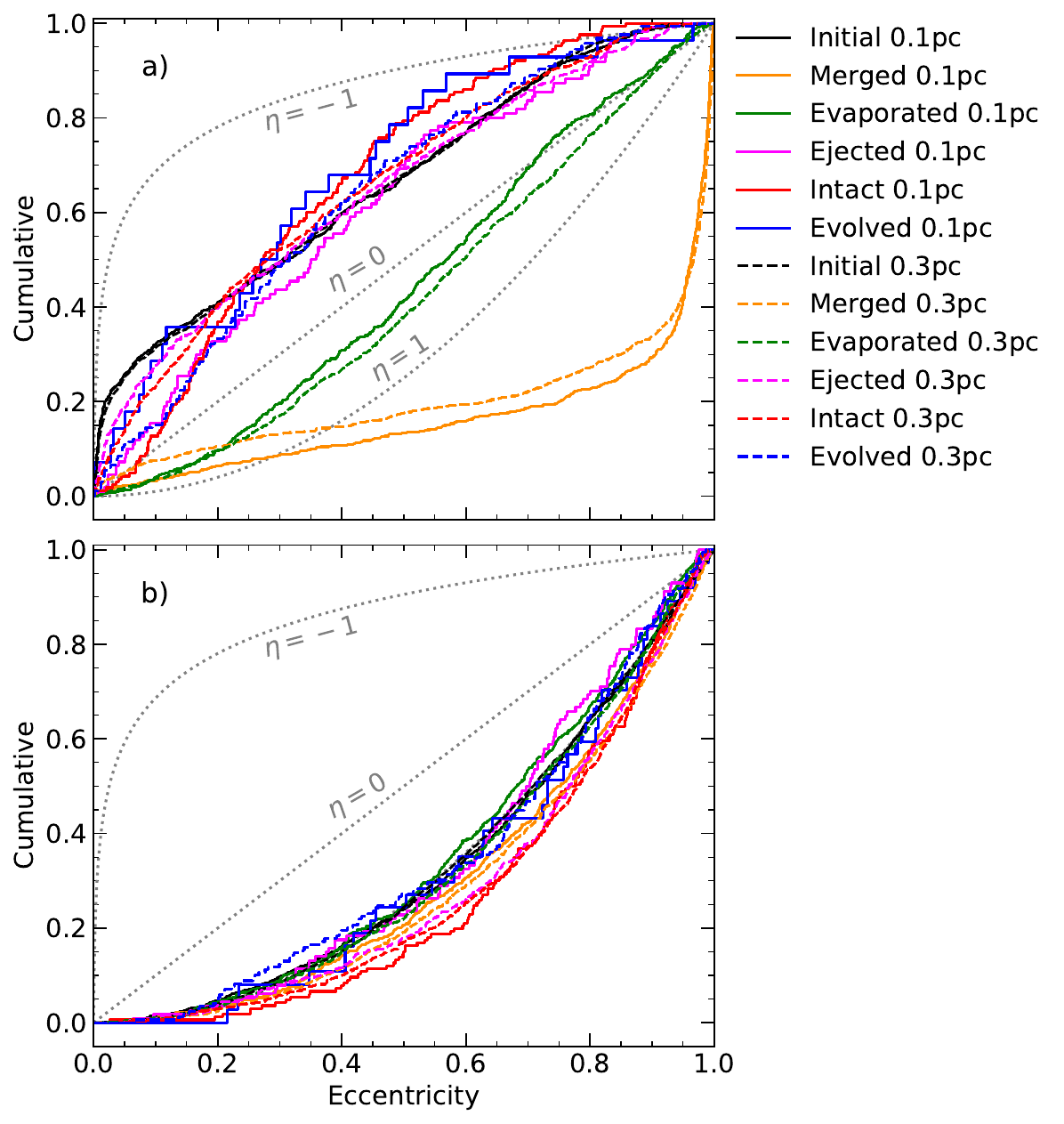}
    \caption{Cumulative distribution of the eccentricities for the inner (panel a) and outer (panel b) orbit, colour-coded as in the previous Figures (outcome). The dotted grey lines follow specific power-law distribution $e^\eta$ where $\eta = -1$ correspond to mostly circular orbits, $\eta = 0$ to a uniform distribution and $\eta = 1$ to a thermal distribution (e.g. mostly eccentric orbits). As in Figure \ref{fig: sma distributions}, the coloured distributions consider the final state of a binary where it can still be considered a binary (and not destroyed). Best-fit parameters can be found in Table \ref{table: power law fits}.}
    \label{fig: ecc distr}
\end{figure}

Merger is the most common outcome for binaries starting with $a_\bullet=0.1\,{\rm pc}$ ($46\%$) and approximately equally as common as evaporation at $a_\bullet=0.3\,{\rm pc}$ ($29\%$)\footnote{The percentages refer to the binaries initially at $0.1$ and $0.3~\mathrm{pc}$ separately. In other words, it is not the percentage of the total number of binaries considered (5736), but instead the 1654 and 4082 binaries for each initial distance, respectively.}. This is perhaps the most clear in Figure \ref{fig: evolution1}, which shows the cumulative fraction as a function of time for the various outcomes. The orange boxes in Figure \ref{fig: sma distributions} show the distributions of the semi-major axes of merging binaries just before the merger.  It can be seen that the merging process is not dominated by hardening of the binaries -- the most common semi-major axis for merger is in both cases similar to the initial most common value around $1\,{\rm au}$. Although a fraction of mergers occur in very tight binaries in which the semi-major axis $a \sim R_{1,2}$, where $R_{1,2}$ is the sum of the stars' radii, the majority of the merging binaries are much wider. More explicitly, it is not a decrease in separation that drives mergers, but rather a rapid increase in the binary's eccentricity. The cumulative distribution in eccentricity for the different outcomes in Fig.~\ref{fig: ecc distr} shows this quite clearly. The initial distribution is weighted towards more circular orbits with a median initial eccentricity of around one-third. The corresponding best-fit eccentricity distribution $P(e)\propto e^\eta$ has $\eta < 0$. However, the mergers are strongly superthermal, with only around a third of merging binaries having $e<0.9$ (best fits for the eccentricity distributions of all outcomes are tabulated in Table \ref{table: power law fits}).

\begin{figure}
    \centering
    \includegraphics[width=\linewidth,trim=0cm 0cm 0cm 1cm, clip]{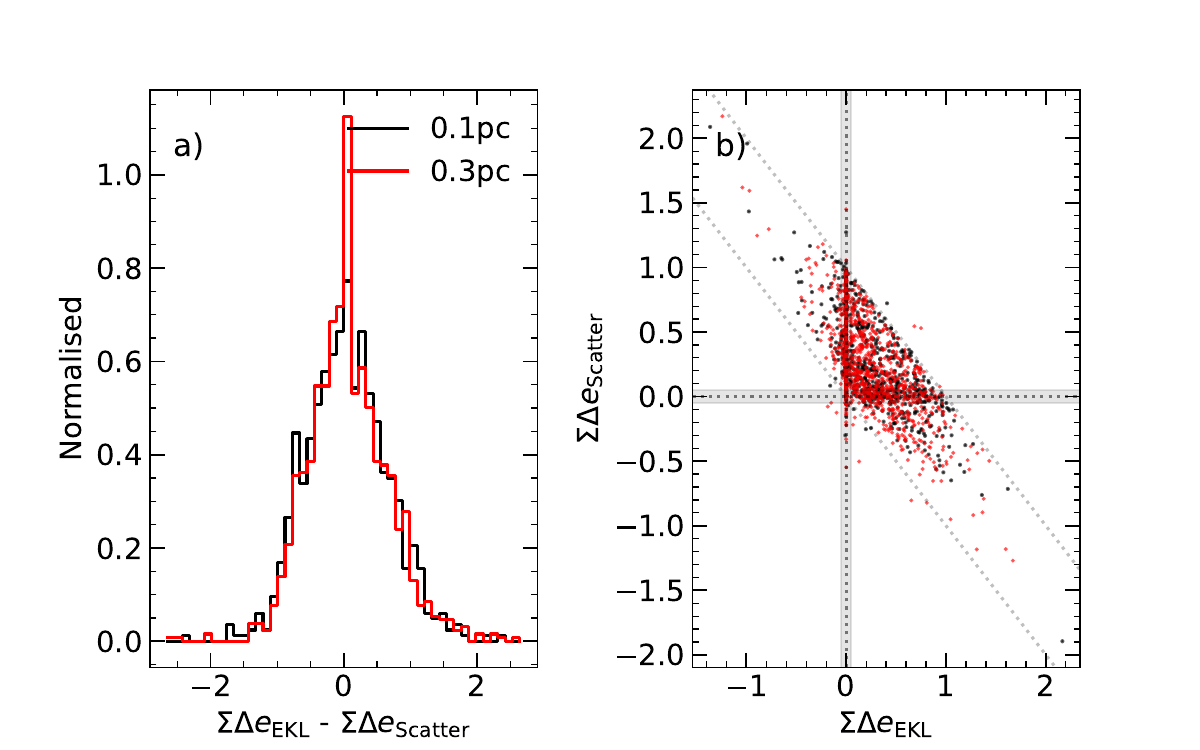}
    \caption{{\it Panel a:} The distribution of the differences between changes in eccentricity due to ZLK and due to repeated scatters (only looking at binaries that eventually merge). The black line corresponds to binaries initially $0.1~\mathrm{pc}$ away from the SMBH; the red line to $0.3~\mathrm{pc}$. {\it Panel b:} Scatter of these two terms; one point corresponds to one binary. The zero-lines are shown as dotted black lines, and the shaded areas correspond to the $\pm 0.05$ region around the zero-lines. The two diagonal lines show $x+y=0$ (lower) and $x+y=1$ (upper).}
    \label{fig: diff ZLK Scat}
\end{figure}

\subsection{Mergers} \label{sect:Mergers}

\begin{table}
    \centering
    \caption{Final outcomes, numbers and fractions at $0.1~\mathrm{pc}$ and $0.3~\mathrm{pc}$.}
    \begin{tabular}{c|cc}
         Outcome &  N ($0.1~\mathrm{pc}$) & N ($0.3~\mathrm{pc}$) \\ \hline
         Merged &  755 \REV{(46\%)} & 1180 \REV{(29\%)}\\ 
         Evaporated & 575 \REV{(35\%)} & 1235 \REV{(30\%)}\\ 
         Evolved & 37 \REV{(2\%) }& 169 \REV{(4\%) }\\ 
         Intact & 166 \REV{(10\%)} & 924 \REV{(23\%)} \\ 
         Ejected & 114 \REV{(7\%) }& 571 \REV{(14\%)} \\\hline \hline 
         Total & 1654 & 4082
    \end{tabular}
    \label{tab: final outcomes}
\end{table}

This result is not unexpected, since the binaries suffer ZLK oscillations which drives the eccentricity to values near unity \citep{Prodan+2015,2016Stephen&Naoz, StephanNaoz2019}. However, we also find that on average, repeated scatters contribute about as much as ZLK does in driving binaries to merge.
The left-hand side panel of Figure \ref{fig: diff ZLK Scat} shows a normalised histogram of the difference between the total change in eccentricity due to ZLK and that due to repeated scatters for merging binaries. That is, for a given binary, we compute the total change in eccentricity solely caused by ZLK, and then subtract the total change solely caused by scatters. The result indicates that ZLK and scatters on average contribute equally in driving up the binary's eccentricity until it eventually merges. The right-hand side panel instead shows a scatter plot of these two terms; ZLK along the x-axis and repeated scatters along the y-axis. A typical binary will increase its eccentricity through both processes. In some cases, however, the two processes work against each other; one process causes a net decrease (the binary becomes more circular) while the other causes a correspondingly larger net increase.

\begin{figure}
    \centering
    \includegraphics[width=\linewidth, trim={0 0cm 0 1.2cm},clip]{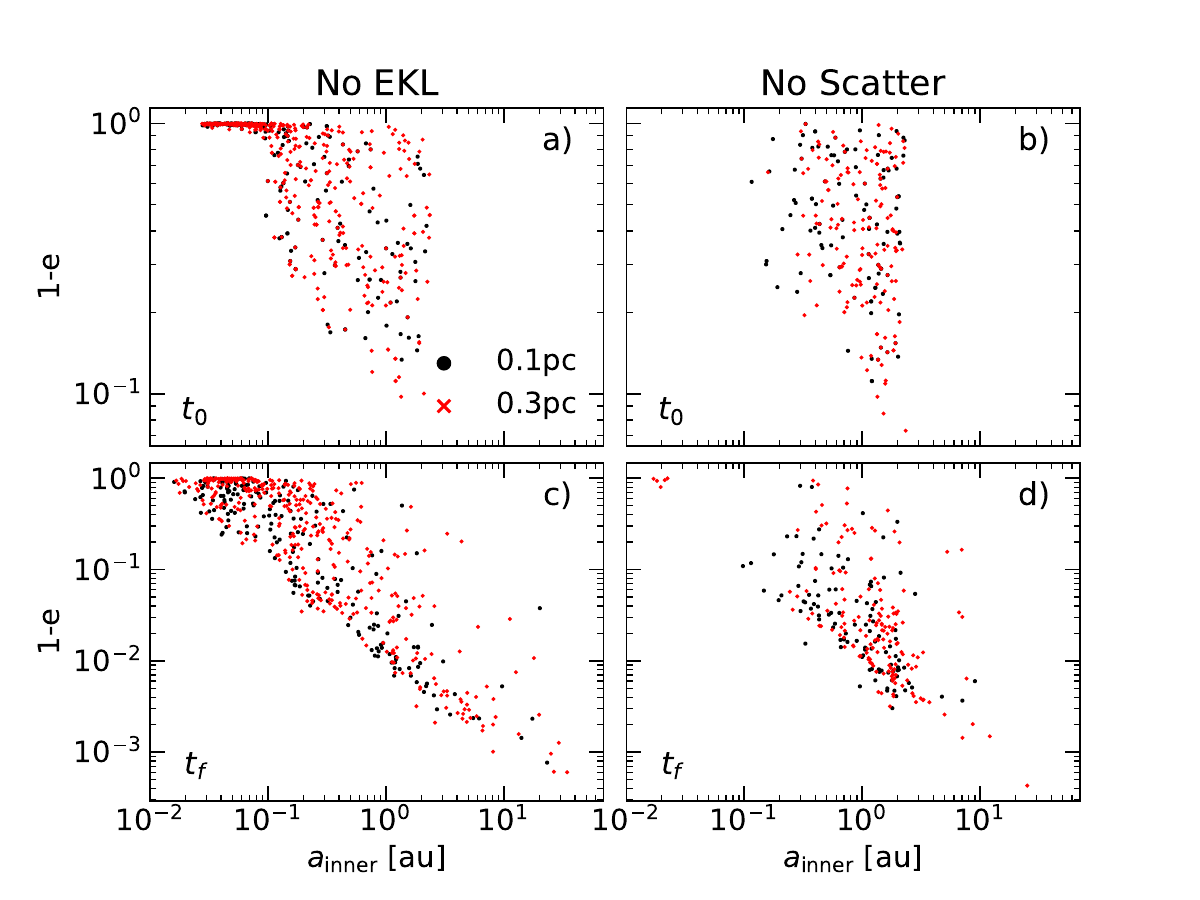}
    \caption{Four-panel plot all showing the separation versus $1-e$ of the inner orbit. Based on the zero-lines in Figure \ref{fig: diff ZLK Scat}; the left-hand side column corresponds to the vertical zero-line ($|\Sigma \Delta e_\mathrm{ZLK}| \leq 0.05$), and the right-hand side column corresponds to the horizontal zero-line ($|\Sigma \Delta e_\mathrm{Scatter}| \leq 0.05$). The top row shows the initial distribution at $t=0$ while the bottom row shows the final distribution at $t=t_f$ before the binaries merge.}
    \label{fig: dominates ZLK Scat}
\end{figure}

We also see some clustering around the zero lines from the right-hand side panel; these are binaries that on average see almost no change from one of the processes and, as such, are dominated by the other. Figure \ref{fig: dominates ZLK Scat} shows the initial (top row) and final (bottom row) separation versus $1-e$ for these binaries. For scatter-dominated mergers ($|\Sigma \Delta e_\mathrm{ZLK}| \leq 0.05$ - shaded area along the vertical zero line), the binaries are initially quite hard and circular. Although many of these binaries become softer and more eccentric, the bulk of them remain hard and relatively circular. They are not undergoing ZLK oscillations because GRP suppresses ZLK for hard and tight binaries; as such, their mergers are driven only by scattering.

\begin{figure}
    \centering
    \includegraphics[width=\linewidth]{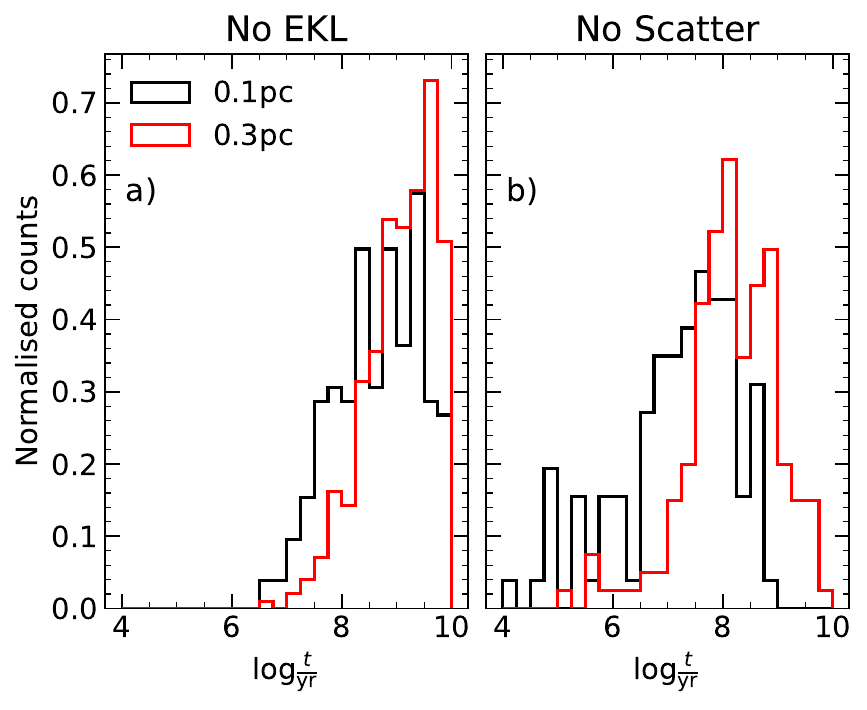}
    \caption{Time of merger distribution for the binaries in Figure \ref{fig: dominates ZLK Scat}. Panel \textit{a)} is, as before, referring to the vertical zero-line (effectively no ZLK) in Figure \ref{fig: diff ZLK Scat} while panel \textit{b)} is referring to the horizontal zero-line (effectively no scatters). Black (red) line is for binaries initially at $0.1~\mathrm{pc}$ ($0.3~\mathrm{pc}$) away from the SMBH.}
    \label{fig: ZLK v Scat times}
\end{figure}

Conversely, the ZLK dominated mergers ($|\Sigma \Delta e_\mathrm{Scatter}| \leq 0.05$ - shaded area along the horizontal zero line) are initially quite wide and soft with moderate eccentric orbits and also remain wide and soft, though naturally become more eccentric due to ZLK oscillations. Wide binaries should (and do) suffer scatters at a higher rate than tight binaries, but the timescale at which these scatters induce a merger is typically an order of magnitude longer than ZLK. Figure \ref{fig: ZLK v Scat times} shows the time distribution of when the mergers shown in Figure \ref{fig: dominates ZLK Scat} occur. On average, a scatter-dominated evolution (left-hand side panel) leads to mergers after $10^9$ years, while for a ZLK dominated evolution it only takes $< 10^8$ years.

\begin{figure*}
    \centering
    \sidecaption
    \includegraphics[width=12cm]{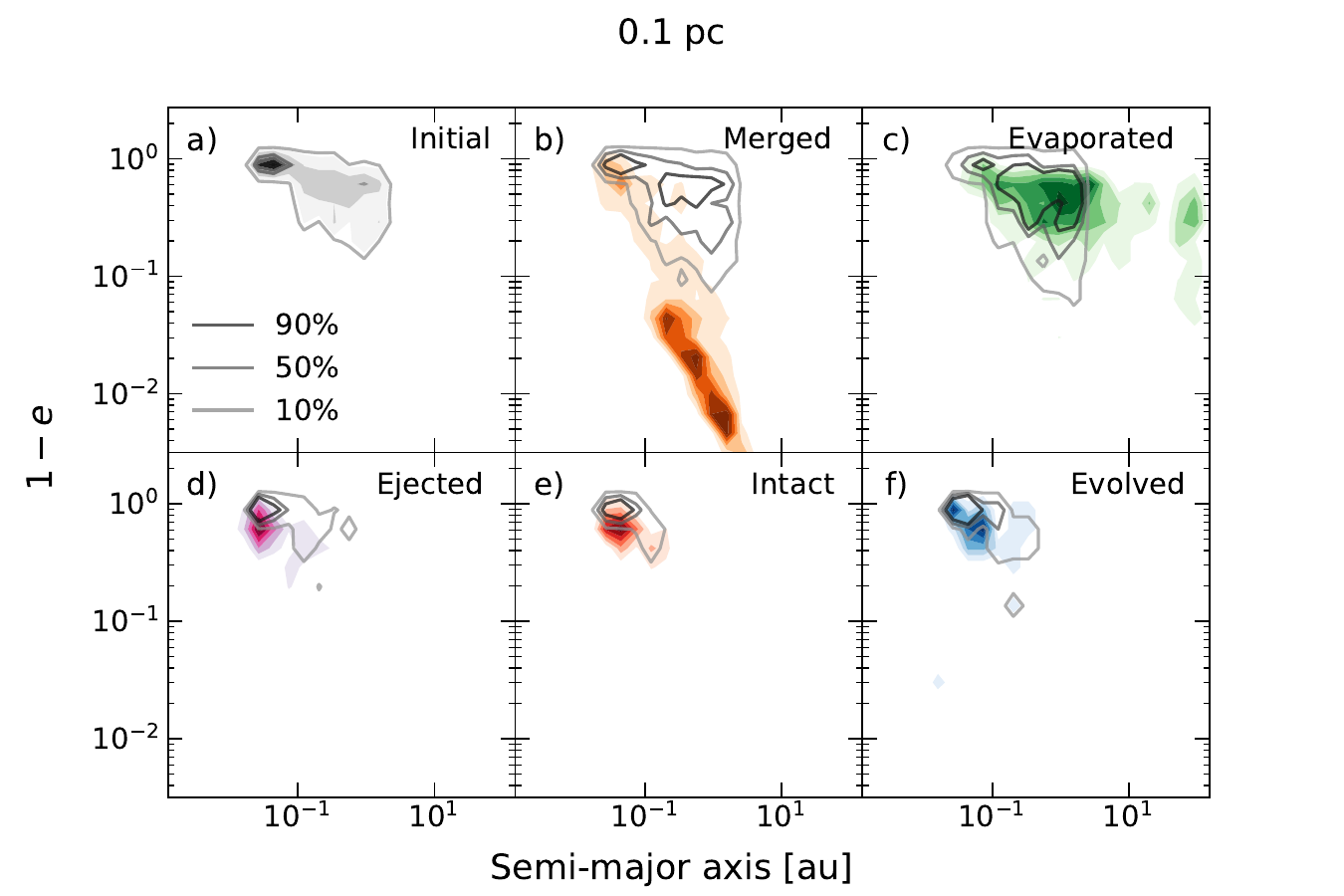}
    \caption{Contours of the inner semi-major axis versus $1-e$ for various outcomes at $0.1~\mathrm{pc}$ from the SMBH. Upper row from the left: initial distribution in black (a), merged binaries in orange (b) and evaporated binaries in green (c). Bottom row from the left: the intact population (after 10 Gyr) in red (d), the evolved population in blue (e) and finally ejected binaries in magenta (f). For panels b), c) and d), the distribution refers to the binary's properties the moment just before the final outcome.}
    \label{fig: sep ecc 1}
\end{figure*}

\begin{figure*}
    \sidecaption
    \centering
    \includegraphics[width=12cm]{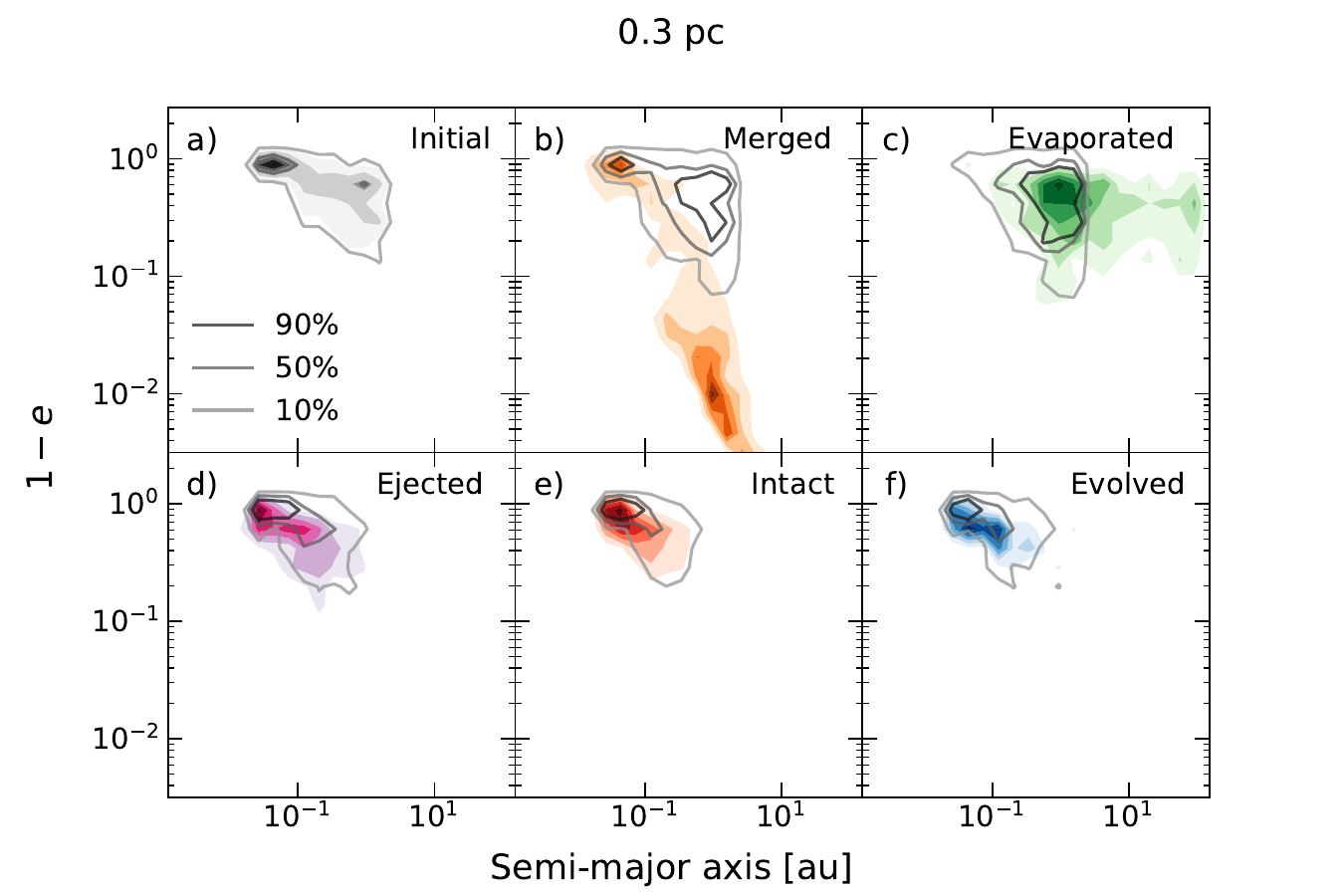}
    \caption{Same as Figure \ref{fig: sep ecc 1} but at $0.3~\mathrm{pc}$.}
    \label{fig: sep ecc 2}
\end{figure*}

To demonstrate that there are two distinct groups of mergers, we plot the two-dimensional distributions of the inner semimajor axis $a$ and eccentricity ($1-e$) grouped by outcome in  Figures~\ref{fig: sep ecc 1}~and~\ref{fig: sep ecc 2}. 

In panel a) the total initial distribution is shown in black, mergers are shown in orange in panel b), evaporated binaries are shown in green in panel c), and the remaining panels d-f in the bottom row, show ejected binaries in magenta, intact binaries in red, and evolved binaries in blue, respectively. Each panel also has three contours (10\%, 50\%, and 90\% of the maximum number of counts) corresponding to the initial distribution for the binaries in that panel. The panel for mergers shows two distinct regions. The first, with about two thirds of mergers, are the initially wide binaries driven by an increase to the their eccentricities mostly due to ZLK and merge at very high eccentricities. The second is the merger of initially tight binaries, driven mainly by scattering, which then merge over a wide range of eccentricities. 

\subsection{Evaporation}
\label{sect:Evaporation}
Evaporation (green lines in Figures~\ref{fig: sma distributions}~to~\ref{fig: ecc distr}) is slightly more common than merger at $0.3\,{\rm pc}$ ($30\%$) and somewhat less common at 0.1\,pc ($35\%$). In particular, evaporation dominates for wider inner orbits $\gtrsim 1~\mathrm{au}$\footnote{Binaries wider than about $0.05-0.15~\mathrm{au}$ are soft between $0.1-0.3~\mathrm{pc}$.}; beyond $\sim 100~\mathrm{au}$ the binary unbinds as it exceeds its Hill radius due to the SMBH. For smaller initial separations, it is predominantly scatters that unbind the binary, either via a strong encounter that ionises it or through repeated weaker encounters that gradually inject energy. The green contours in Figures \ref{fig: sep ecc 1} \& \ref{fig: sep ecc 2} (panel c) show the distribution of binaries just before they evaporate; the smaller the semi-major axis is the stronger the encounter that causes the binary to unbind needs to be. In other words, if a binary evaporates at $0.1~\mathrm{au}$, it is ionised by a very strong encounter, and if it evaporates at, for example, $ 10~\mathrm{au}$, then it is largely due to a gradual process.

The corresponding final outer orbits seen in Figure \ref{fig: sma distributions} are centered around the initial distances ($0.1,\, 0.3$ $\mathrm{pc}$) and are distributed rather symmetric in log space. The ratio of evaporated and merged binaries is roughly constant throughout the simulations, as seen in Figure \ref{fig: evolution1}. Because both processes are fast, most binaries will therefore not have had time to migrate in either direction before being destroyed, which explains the shape of the distribution in Figure \ref{fig: sma distributions}.\footnote{It is possible this could have effects on the mass segregation in the early stages of the evolution of the NSC assuming there was a substantial binary fraction. However, since the binaries are destroyed so quickly, we would not expect long-lasting effects.} For the inner orbits, evaporations follow a very similar distribution to the merged binaries, but the distribution extends to very wide separations. 
The evaporating binaries have an eccentricity distribution that is close to thermal ($\eta = 1$).  Eccentric binaries are more vulnerable to being broken up than circular binaries, since they spend much of their time at apoapsis, where a smaller velocity perturbation is sufficient to unbind them.

\begin{figure}
    \centering
    \includegraphics[width=\linewidth, trim= 0cm 0.2cm 0cm 1cm, clip]{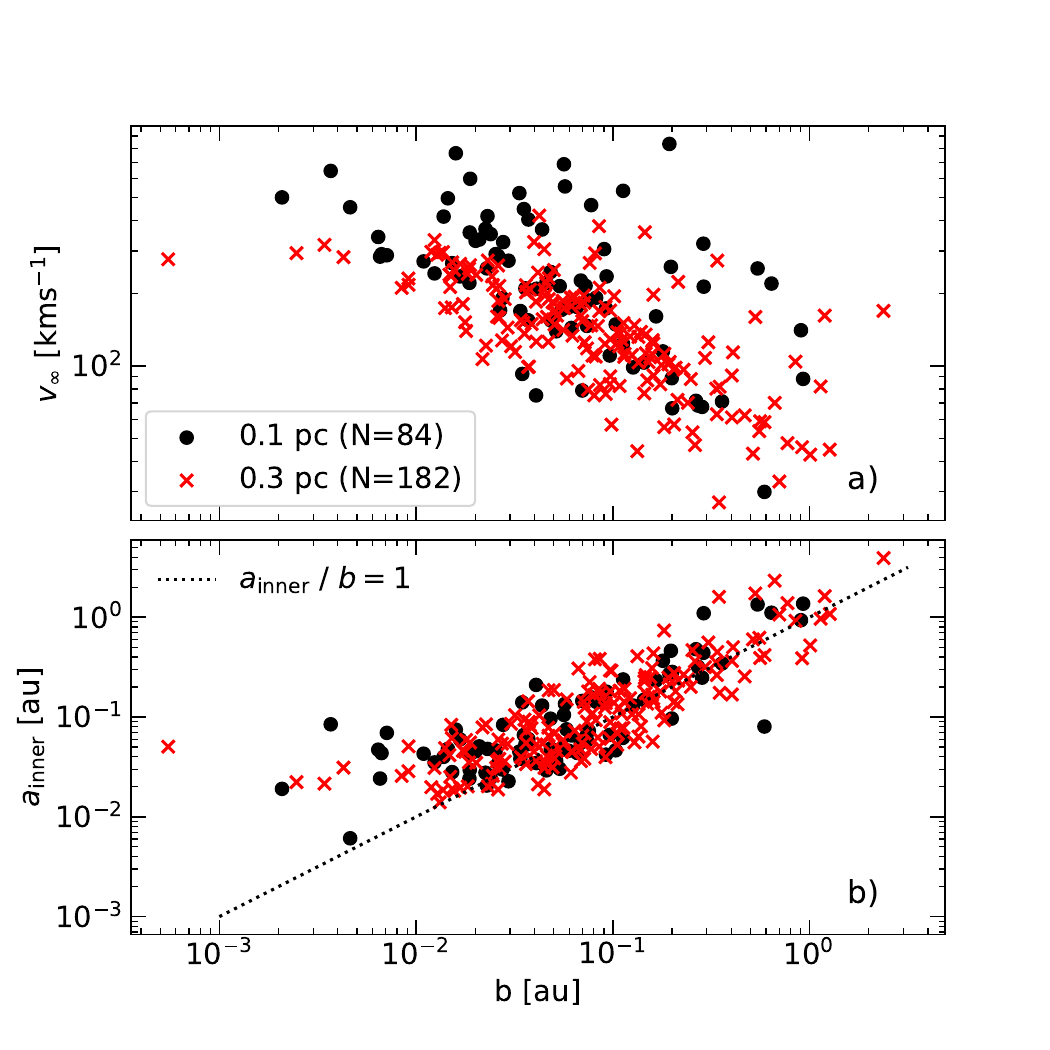}
    \caption{Panel \textit{a)} shows the relation between the tertiary's velocity at infinity and its impact parameter. The points are colour-coded based on their initial position from the SMBH. Panel \textit{b)} shows the impact parameter versus the binary's semi-major axis before the collision. To help guide the eye, the dotted black line corresponds to $a_\mathrm{inner} = b$.}
    \label{fig: b vinf}
\end{figure}

\begin{figure}
    \centering
    \includegraphics[width=\linewidth]{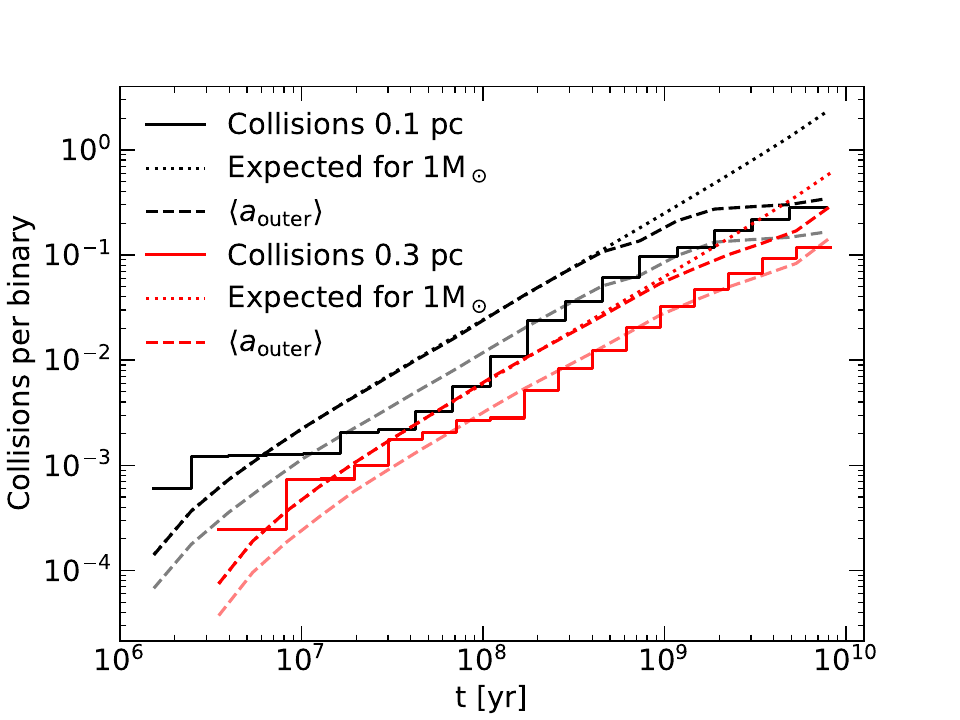}
    \caption{The average number of collisions per binary as a function of time (solid lines). The dotted lines refer to the expected number of collisions a single $1~\mathrm{M}_\odot$ star would have at $0.1~\mathrm{pc}$ (black) and $0.3~\mathrm{pc}$ (red). Because the binaries tend to migrate, the dashed lines show the same expectation as the dotted line, but where the distance is not fixed to the starting positions and instead follow the average outer orbit of the binaries. The pale dashed lines follow the apoapsis, while the non-transparent dashed lines follow the semi-major axis.}
    \label{fig: colls per binary}
\end{figure}

\subsection{Hard binaries} \label{sect:Hard binaries}

Binaries that have not merged or evaporated fall into three categories: ejected binaries (magenta in Figs.~\ref{fig: sma distributions} and \ref{fig: ecc distr}), intact binaries (red) and evolved binaries (blue). All three groups have similarly distributed inner semi-major axes, eccentricities, and are relatively hard; $a<1\,{\rm au}$ but peaks at $\lesssim 0.1~\mathrm{au}$.

The ejection of a binary occurs when it becomes unbound to the SMBH. We do not attempt to follow the subsequent dynamical evolution of such binaries, although they will likely remain within the NSC on wider galactic orbits. On these orbits the number densities are smaller, so they do not undergo significant further close encounters. \REV{Furthermore, GRP should inhibit ZLK oscillations even for the wider binaries (see eqn. \ref{t_GR vs t_quad}). Mass precession, induced by the extended potential of the cluster, may also suppress ZLK; though only at the octupole order\footnote{The octupole-order equations of motion, apart from inducing changes in the inner orbit, also induce changes in the outer orbit, although at longer timescales. It furthermore requires a non-circular outer orbit, and a mass ratio of $q > 1$ for the binary. Even in the most extreme scenario for one of our binaries, e.g. $m_1=1~\mathrm{M}_\odot$, $m_2=0.1~\mathrm{M}_\odot$, $a_\mathrm{inner}\sim3~\mathrm{au}$, the associated timescale for the octupole order is at least $\gtrsim 2-3$ orders of magnitude longer than the quadrupole timescale at distances $\geq 0.1~\mathrm{pc}$ \citep[see][]{Naoz+2013a, Naoz2016}.} as the quadrupole order is independent of the outer argument of periapsis \citep[see e.g.][]{Naoz+2013a,2016Stephen&Naoz,Tep+21}.} We therefore expect that these binaries will either evolve, if their primary star is more massive than the turn-off mass at the end of the simulation, or remain intact.

Similarly, the intact binaries are all found to have migrated away from the SMBH. This is counter-intuitive at first sight, since binaries, which on average are more massive than single stars, preferentially migrate inward due to dynamical mass segregation. However, binaries that migrate inward are consistently destroyed, whereas those that migrate outward are not; the quadrupole timescale becomes longer and the interaction rate with surrounding stars decreases (see, eqn. \ref{eqn: tenc}). \REV{Moreover, and partly for the same reason, surviving binaries are preferentially hard and circular (\newREV{see dashed lines in Figure \ref{fig:h-distr}; virtually all binaries have $h\ge1$.} See also Figures \ref{fig: sma distributions}, \ref{fig: ecc distr}, \ref{fig: sep ecc 1} and \ref{fig: sep ecc 2}). \newREV{In comparison}, three-body encounters involving hard binaries \newREV{in globular clusters}, would typically lead to a thermal or \newREV{super-thermal} distribution in eccentricity \citep{Heggie1975,Heggie1975_Clusters,Stone_Leigh2019,Ginat_Perets2023,RandoForastier+2025}. Coupled with the migration discussed above, very hard and circular binaries do not undergo ZLK nor do they interact much with surrounding stars, and the ones that do get eccentric likely merge or circularise as a consequence. Since the hardness of a binary is a relative term (eqn. \ref{Eq: hardness ratio}), a hard binary in the NSC will have a very small separation\footnote{The hardness is essentially set by the velocity dispersion of the system. For the NSC $\sigma\gtrsim100\mathrm{kms}^{-1}$ while for a globular cluster $\sigma\lesssim10\mathrm{kms}^{-1}$ \citep{Ivanova+2005}, which means that the softest hard binary in a globular cluster needs to have a separation that is at least $\sim100$ times smaller to be considered hard in the NSC.} which further limits the possible eccentricities that do not lead to a merger or the circularisation of the binary. For example, the most common separation for our binaries at $0.1~\mathrm{pc}$ is roughly $0.05~\mathrm{au} \approx 10\mathrm{R}_\odot$ (see figure \ref{fig: sma distributions}), which means that for an eccentricity $>0.8$ the binary is likely to merge and otherwise likely to circularise \citep{Mardling&Aarseth2001}. As such, we do not expect a thermal distribution for old primordial binaries.}

The evolved population is essentially just a subset of the intact population, for a few reasons. In these simulations we follow the evolution of sub-solar mass stars, meaning all of them are on the MS up until the very end of the simulation ($\gtrsim 9~\mathrm{Gyr}$ as seen in Figure \ref{fig: evolution1}). We do not compute any complex stellar evolution during the simulation (see details on stellar evolution in Section \ref{sect: Stellar Populations}), and as such this distinction between evolved and intact binaries is essentially just separating the most massive stars ($\gtrsim 0.97~\mathrm{M}_\odot$) from the rest. Since the majority of binaries are destroyed within the first few billion years, we would not expect a substantial difference between these two populations. The main thing that separates the evolved population from the intact population is its small fraction of binaries that have evolved prematurely ($\lesssim 9~\mathrm{Gyr}$) due to stellar collisions. 

\subsection{Collisions}

With the high number densities of stars, stellar collisions are relatively common in the NSC. Here we present the sample of collisions from our simulation in more detail.

The upper panel of Fig.~\ref{fig: b vinf} shows the impact parameter $b$ versus the velocity at infinity $v_\infty$ for the tertiary in cases where it leads to a collision. The $v_\infty$ velocities range from a few tens to a few hundreds of $\mathrm{kms}^{-1}$. As such, none of these collisions should be able to strip the binary star with which it collides \citep{Rose+2023}. Therefore, treating collisions with no mass loss and conservation of momentum should be a reasonable approximation (see Section \ref{Stellar Interactions}). The lower panel of the Figure shows the relation between the impact parameter and the binary separation for the collisions; we find a scatter around the 1:1 ratio. This is unsurprising since the impact parameter is defined as the distance from the binary COM, $b \sim a$ roughly corresponds to where one of the binary stars is expected to be.

\begin{figure}
    \centering
    \includegraphics[width=0.95\linewidth]{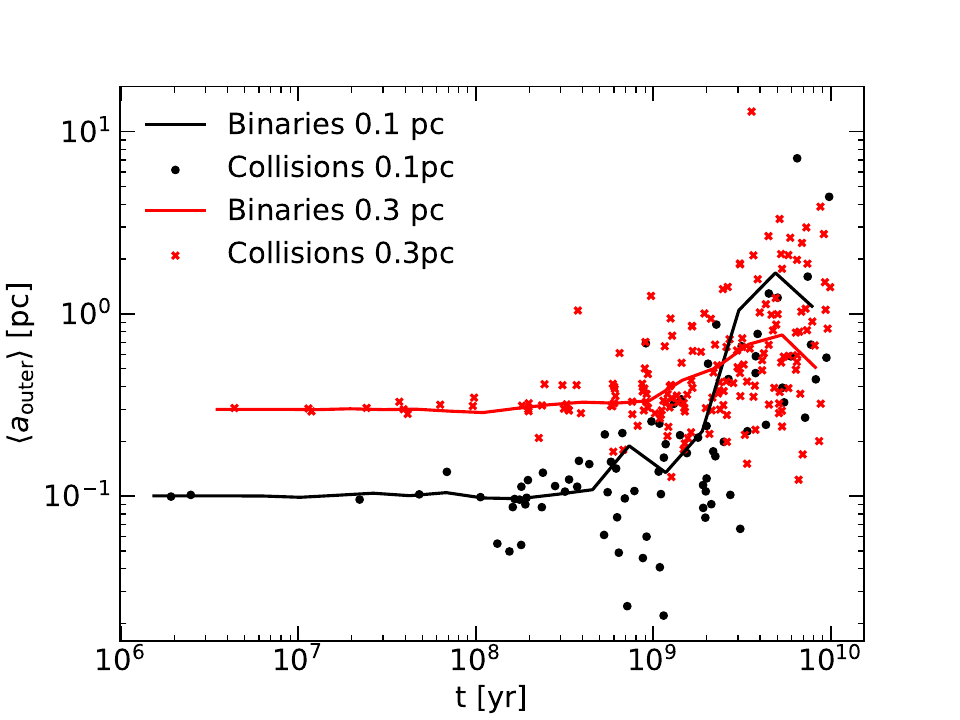}
    \caption{Mean outer semi-major axis for all remaining binaries as a function of time.  Black lines show binaries starting at  $0.1~\mathrm{pc}$ from the SMBH; red lines binaries starting at $0.3~\mathrm{pc}$.  Each point shows the time and location of a collision.}
    \label{fig: colls aouter}
\end{figure}

A consequence of the binaries' tendency to migrate inwards or outwards from the SMBH, is that the collision rate changes significantly compared to a fixed orbit around the SMBH. The solid lines in Fig.~\ref{fig: colls per binary} show the average number of collisions per binary as a function of time. The dotted lines show the expected number of collisions for a single $1~\mathrm{M}_\odot$ star at a fixed galactocentric radius - e.g. after 10 Gyr we would expect a $1~\mathrm{M}_\odot$ star to have suffered at least one collision at $0.3~\mathrm{pc}$ and between 2-3 at $0.1~\mathrm{pc}$. We are almost always an order of magnitude below this expectation, which is caused by the binaries' migration (e.g. the dynamical recoil from three-body encounters, see Section \ref{sect: Evolving a binary}). The dashed lines show the same expectation as the dotted lines, but where the distance to the SMBH changes as the average of the binaries' outer orbits (normal black/red for semi-major axis, and pale black/red for apoapsis). The way in which the average outer orbit varies as a function of time is shown in Figure \ref{fig: colls aouter}. After a few Gyr, the yet intact binaries have on average migrated well beyond their starting points, and as a consequence most collisions with a binary (which there are less and less of) also occur further out in the NSC. The transparent dashed lines in Figure \ref{fig: colls per binary}, which follow the average apoapsis of the outer orbits, is the only expectation that translates to the actual rate of collisions per binary; binaries spend most of their \newREV{(outer)} orbits at apoapsis, leading to a much lower collision rate.

\begin{figure}
    \centering
    \includegraphics[width=0.95\linewidth]{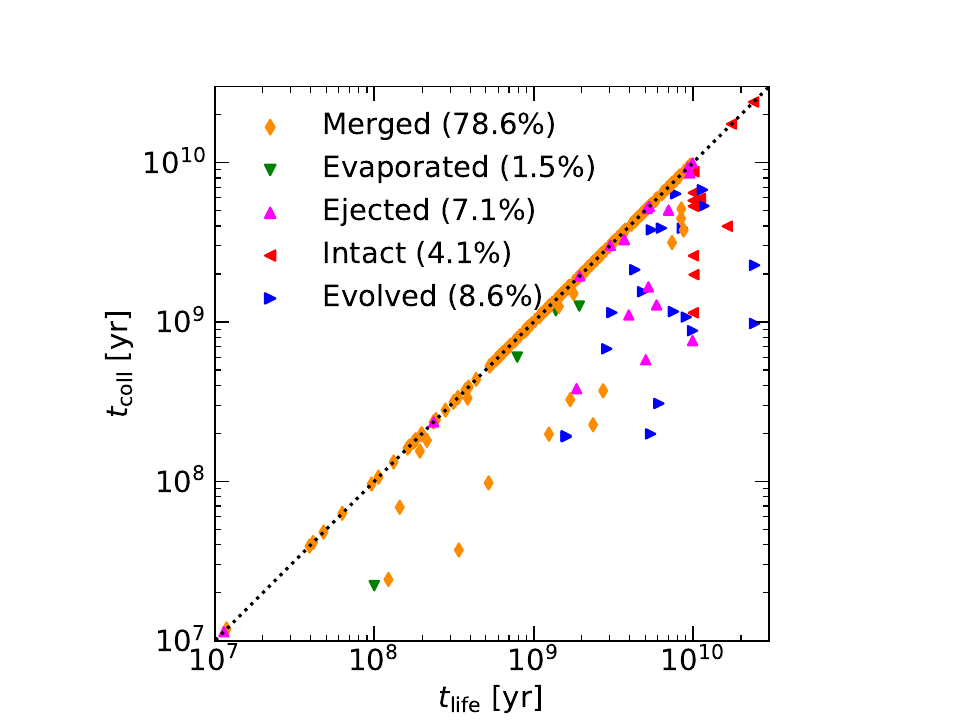}
    \caption{Time of final outcome versus the time of collision for binaries that have suffered at least one collision. The dotted black line shows where $t_\mathrm{life} = t_\mathrm{coll}$; points on this line meet their final outcome as a direct consequence of the collision. The points are colour-coded based on the outcome similarly to Fig. \ref{fig: sma distributions}.}
    \label{fig: tlife v tcoll}
\end{figure}

Figure \ref{fig: tlife v tcoll} shows the time of the final outcome $t_\mathrm{life}$ of binaries that have suffered at least one collision versus the time of collision $t_\mathrm{coll}$. The black dotted line marks where $t_\mathrm{life} = t_\mathrm{coll}$; for a binary on this line the outcome is a direct consequence of the collision. The percentage for each outcome is also indicated in the legend of the Figure; most commonly, a collision leads to a merger between the remaining stars, and in most cases this also happens as a direct consequence of the collision. We refer to these mergers as {\it three-body pile-ups} (3BPUs).

The second most common outcome is the intact + evolved + ejected outcomes, which are covered in Section \ref{sect:Hard binaries}. These collisions have had little effect on the subsequent evolution of the binaries and so the final outcome is usually significantly later than the collision. The same goes for the small fraction of binaries that evaporate - while they are not a direct consequence of the collision, they tend to happen very shortly afterward.

\section{Discussion}\label{Discussion}

In the previous section, we give an overview of the many different outcomes a binary in the NSC may suffer, and the various processes that inevitably steer the binary's evolution towards one of the outcomes. 

The parameter space that we consider for the binaries in these simulations is quite stringent, which makes the varied distribution of outcomes interesting; the path a binary takes seems to be largely stochastic. While processes like ZLK oscillations are well behaved (in the sense that you can predict the oscillations), encounters with surrounding stars are not. The consequences of this are particularly apparent for binaries at (or near) the H/SB. 
For example, as seen from Figure \ref{fig: evolution1}, at $0.3~\mathrm{pc}$ a binary (from our sample) is essentially as likely to remain intact as it is to merge or evaporate. Figures \ref{fig: sep ecc 1} and \ref{fig: sep ecc 2} also show that the initial distribution in separation and eccentricity is very similar between the outcomes; the contours are virtually the same for binaries that eventually merge and evaporate. The 90\% contours, which show two distinct regions in both cases, only differ in extent between the two outcomes. Evaporated binaries are shifted towards greater separations (these are soft and not near the hard/soft limit) while the merged binaries are shifted towards smaller separations and look virtually identical to binaries that remain intact, evolve or get ejected. As discussed in Section \ref{sect:Hard binaries}, the only real difference between the intact and evolved binaries is the mass, but in the end, the outcome will still be the same. Although unpredictable, binaries that are moderately soft to soft tend to either merge or evaporate (latter is more likely for softer binaries), while binaries that are moderately soft to moderately hard either merge, remain intact (evolve) or get ejected. Which one seems to be largely up to chance.

In the following section(s) we narrow the discussion down to binary mergers, and connect it to recent publications and observations of stars, binaries and other peculiar objects in the NSC.

\subsection{Binary mergers and the S-cluster} \label{Discussion: Binary mergers}

Binary mergers are perhaps the most interesting aspect of the outcomes in a binary's evolution in the NSC. The leading theory for the origin of the G-objects \citep{Ciurlo+2020} is that they form as a consequence of binary mergers \citep{Prodan+2015,2016Stephen&Naoz,StephanNaoz2019,Ciurlo+2020,Pei_ker_2024}, and that they may be related to the lower mass population of the S-cluster \citep{melamed2024updatedustysourcescandidate}.

\begin{figure}
    \centering
    \includegraphics[width=\linewidth]{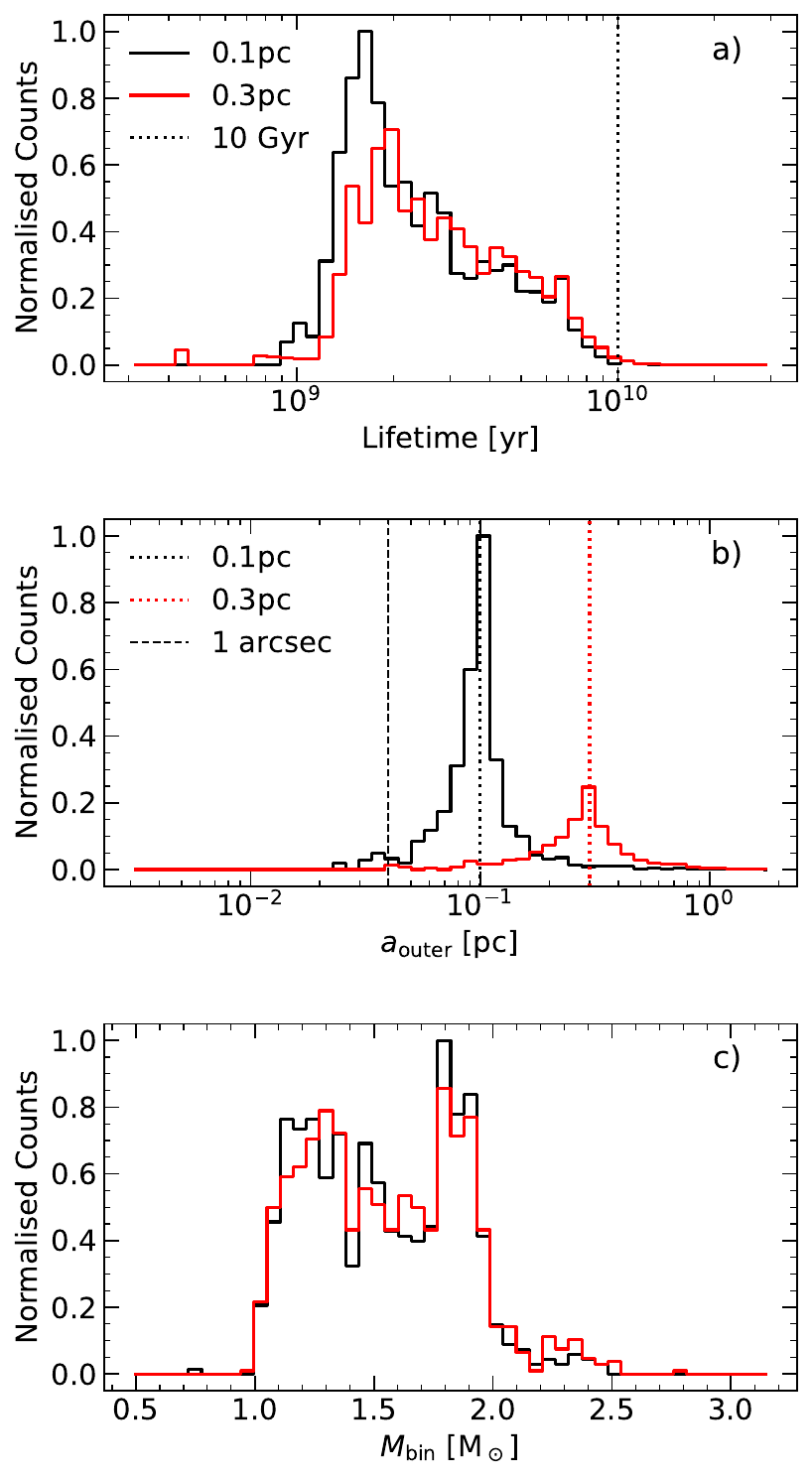}
    \caption{{\it Panel a):} Normalised histograms for the total expected lifetime of the mergers. The dotted black line marks the age of the NSC at 10 Gyr. {\it Panel b):} Post-merger semi-major axis around the SMBH. The dotted vertical lines mark the initial separations. The solid vertical line marks 1 arcsecond. {\it Panel c):} total mass after the merger. For all panels, binaries originating at $0.1~\mathrm{pc}$ ($0.3~\mathrm{pc}$) are shown in black (red).}
    \label{fig: merger orbits hists}
\end{figure}

Figure \ref{fig: merger orbits hists} illustrates various properties of the post-merger binaries. The solid black lines represent mergers of binaries initially located at $0.1~\mathrm{pc}$ from the SMBH, while the solid red lines correspond to those initially positioned at 
$0.3~\mathrm{pc}$. The top panel depicts the distribution of the merger products' total lifetimes; that is, the time spent in the binary plus the additional time gained from the merger ``rejuvenation''\footnote{The merger-product's expected rejuvenated lifetime is calculated using the same core helium mass-fraction prescription as for collisions, see Section \ref{sect: Stellar Populations}.}. The dotted black line marks 10 Gyr - the time considered for our simulations, which roughly corresponds to the age of the NSC. A small fraction ($\sim 1\%$) of merger products are located to the right of this line, suggesting a few mergers originating from the early populations of binaries in the NSC could be observed at present day either as an MS star or as a recently evolved star. Of course, the shape of the distribution(s) is largely due to our choice to focus on binaries where the primary is near the MS turn-off mass of the NSC \citep{Pfuhl+2011,Schodel+2020, NSC_history2023}; considering more massive stars would push the distribution towards the left\footnote{The shortest lifetimes of the merger products can be compared to the turn-off age for a $2~\mathrm{M}_\odot$ solar-metallicity star, which is $\sim 1.1~\mathrm{Gyr}$.}, and less massive stars would push it to the right. In other words, the same mechanism can make massive blue straggler stars (BSSs) at earlier times and less massive ones today.

The middle panel shows the distribution of the merger products' orbits around the SMBH, where the black dotted line and red dotted line correspond to the initial positions for the binaries: $0.1~\mathrm{pc}$ and $0.3~\mathrm{pc}$, respectively. Most of the post-merger orbits remain at these distances, as can be seen from the two distinct peaks. Not surprisingly, the $0.1~\mathrm{pc}$ peak is much narrower; the relevant timescales are shorter, which, apart from causing the binaries to merge more quickly (see e.g. Figure \ref{fig: evolution1}), also means that the binaries do not have a lot of time to migrate (in either direction). Still, there is a small fraction ($\sim 1\%$, all of which originate from $0.1~\mathrm{pc}$) of binaries that have migrated inward towards the SMBH, obtaining an outer semi-major axis smaller than about 1 arcsecond ($\approx 0.04~\mathrm{pc}$), see also Figure \ref{fig: sma distributions}. 

This is roughly the region that contains the orbits of the S-cluster, including the G-objects and dusty sources \citep{Habibi+2017, Schodel+2020, Pei_ker_2024, melamed2024updatedustysourcescandidate}. The results therefore support the idea that the S-cluster and G-objects, may originate further out in the NSC \REV{\citep[see also, for example,][]{Verbene+2025}}. However, in our case, there is a mismatch in timescales and age: the S-cluster and the G-objects appear to be young $\lesssim 10-15~\mathrm{Myr}$ \citep{Habibi+2017, Schodel+2020, Pei_ker_2024, melamed2024updatedustysourcescandidate}, which is comparable (in the context of the G-objects) to the ZLK merger timescale for binaries on orbits $a_\bullet \lesssim 0.1~\mathrm{pc}$ \citep[see e.g.][]{ 2016Stephen&Naoz, StephanNaoz2019}. Compared to our simulations, the binary mergers end up within the inner arcsec after at least $\sim300~\mathrm{Myr}$ with the mean being $660~\mathrm{Myr}$, see Appendix \ref{Appendix: Inner arcsec orbits}. Although there is a degeneracy between stellar age and metallicity for S-cluster stars \citep{generozov2024sstarszoneavoidancegalactic}, it is not strong enough to account for the differences seen here. Furthermore, with the recent observation of a lower-mass spectroscopic binary system in the NSC \citep{Pei_ker_2024}, estimated to be only a few Myr old and most likely a progenitor of the G-objects, another mechanism that is not in tension with the recent star formation in the GC \citep{Lu+2013,Schodel+2020} is required \citep[see for example][]{ Genzel+2010, trani2019a, generozov2024sstarszoneavoidancegalactic, Verbene+2025}.

However, Figure \ref{fig: merger orbits hists} shows that the mergers of binaries originating between $0.1-0.3~\mathrm{pc}$ from the SMBH can contaminate the region where the S-cluster is. Although most stars in the NSC are old ($\gtrsim 5-8~\mathrm{Gyr}$), smaller fractions of stars are associated with star formation events at different epochs \citep{Schodel+2020}. The second most prominent event in terms of the mass fraction formed per Gyr is believed to have occurred $2-4~\mathrm{Gyr}$ ago \citep[see, for example][]{Schodel+2020}. As seen in the top panel of Figure \ref{fig: merger orbits hists}; the peak of the total expected lifetime for the mergers is around $2~\mathrm{Gyr}$ for binaries originating both at $0.1~\mathrm{pc}$ (black) and $0.3~\mathrm{pc}$ (red), suggesting that there could be a population of BSSs at or just beyond the MS turn-off scattered in the NSC (see middle panel) associated with this SFE. 

Assuming the NSC is $\sim 10~\mathrm{Gyr}$ old (dotted black line in the top panel), virtually all mergers from the oldest population of stars in the NSC would have evolved at present day due to single star stellar evolution. \citet{Thorsbro+2023} reported metallicities for three $\sim\mathrm{Gyr}$ old stars in the NSC. One of these, which has an apparent age of $1.5~\mathrm{Gyr}$ and a very low metallicity of [Fe/H] = -1.2, was suggested to be a potential BSS. In their hypothesis (which was not presented as a definite model), this BSS would have formed from an initially hard ($a \sim 6~\mathrm{R}_\odot$) and circular binary with stars of $0.9$ and $0.8~\mathrm{M}_\odot$ - very similar to the binaries considered in our simulations. Following a phase of stable mass transfer at $8.3~\mathrm{Gyr}$, the binary would eventually merge at an age of $9~\mathrm{Gyr}$. In their test to this hypothesis, the binary is evolved in isolation, e.g. not in the environment of the NSC. \REV{Based on \citet{Rose} and their analytical study of binaries in the inner $\lesssim 0.5~\mathrm{pc}$, such a binary could remain intact for roughly that time-period outside $\gtrsim0.3\mathrm{pc}$\footnote{\citet{Thorsbro+2023} do not provide a distance between the star and the SMBH}}. Figures \ref{fig: sep ecc 1} and \ref{fig: sep ecc 2} show that such hard and circular binaries can survive both from $0.1~\mathrm{pc}$ and $0.3~\mathrm{pc}$ for up to $10~\mathrm{Gyr}$ in the NSC despite its harsh environment, thanks to outward migration. \REV{Alternatively, the binary could originally have been both wider and more eccentric, before circularising.} As such, the premise that this star may originate from a binary merger associated with an older population of the NSC seems perfectly reasonable. Additionally, it also suggests that more stars similar to it should be present in the NSC; at present day such stars could be located along the asymptotic giant branch on the HR-diagram \citep[where the star in][was observed]{Thorsbro+2023}.

The bottom panel in Figure \ref{fig: merger orbits hists} shows the distribution of masses for the merger-products. A small fraction of these have masses above $2~\mathrm{M}_\odot$; since the maximum mass for each of our sampled stars are $1~\mathrm{M}_\odot$, the only possible explanation is that an additional star (tertiary) is included via a collision. We find that when a physical collision occurs between a star in a binary and an incoming single star, in 80\% of the cases the binary merges as a consequence. As seen in Fig.~\ref{fig: tlife v tcoll}, most of these mergers are also instantaneous in the sense that they occur as a direct consequence of the collision. These 3BPUs serve as a way to form more massive BSSs. 

\begin{figure}
    \centering
    \includegraphics[width=0.95\linewidth]{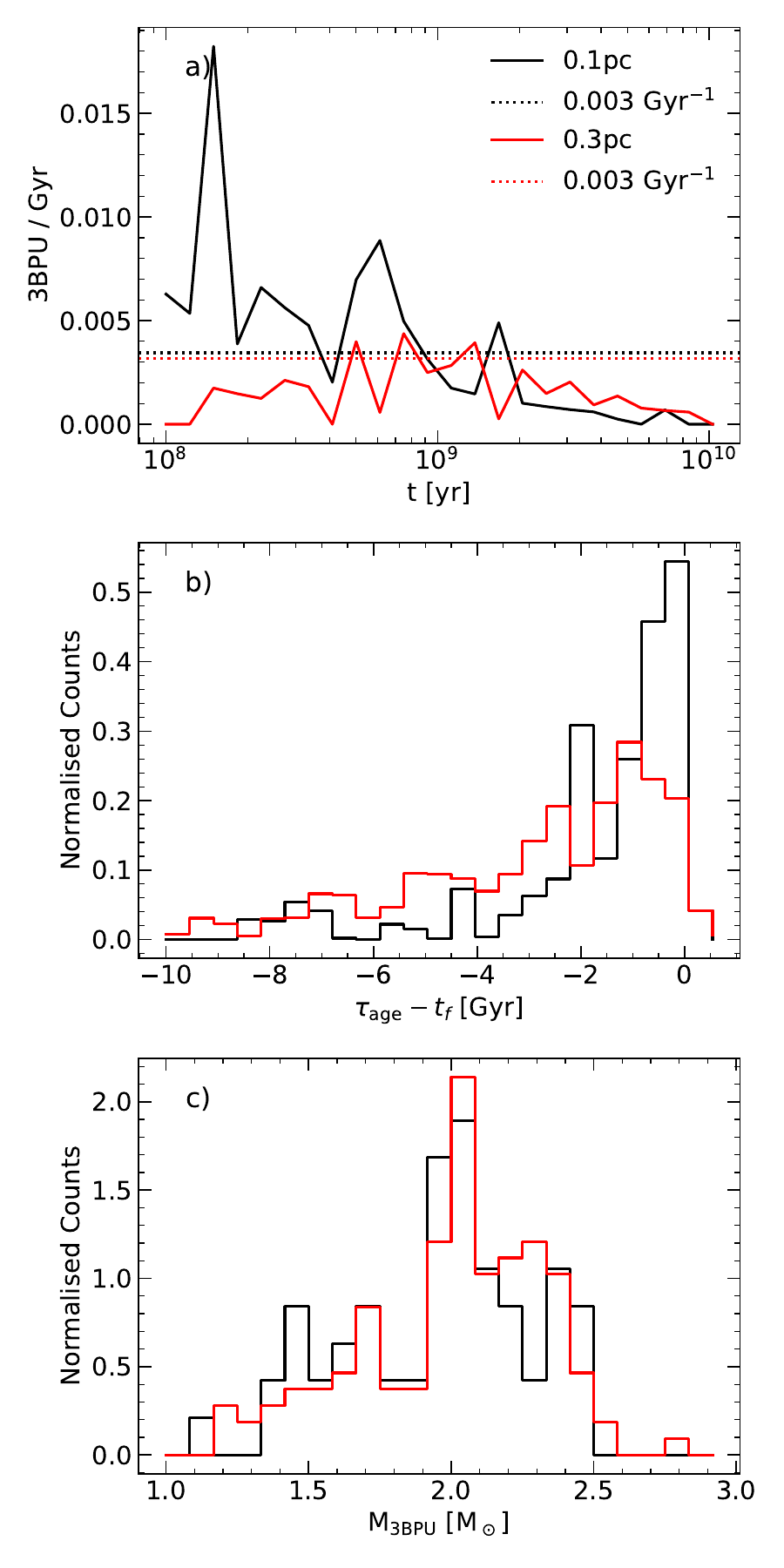}
    \caption{{\it Panel a):} Rate of 3BPUs as a function of time. The dotted lines correspond to the mean rates over a 10 Gyr period. {\it Panel b):} Normalised distribution of the estimated apparent age for the 3BPU merger products. The apparent age is determined by mapping the Helium-core mass fraction of the primary at the time of the merger, to the corresponding age a normal star with the mass of the merger product would be with an identical Helium-core mass fraction. {\it Panel c):} Normalised distributions of the total masses of the 3BPU merger products. For all panels, binaries originating at $0.1~\mathrm{pc}$ are shown in black and those from $0.3~\mathrm{pc}$ in red.}
    \label{fig:3BPUs}
\end{figure}

The top panel of Figure \ref{fig:3BPUs} shows the rate $\frac{N_\mathrm{3BPU}(t)}{N_\mathrm{bin}(t) \times t}$ of 3BPUs, normalised to the number of intact binaries, as a function of time. Although the mean 3BPU rates over 10 Gyr are 0.0034 $\mathrm{Gyr}^{-1}$ and 0.0032 $\mathrm{Gyr}^{-1}$ for binaries originating at $0.1~\mathrm{pc}$ and $0.3~\mathrm{pc}$, respectively, most 3BPUs occur within $\lesssim 1-2~\mathrm{Gyr}$. Given the recent star formation events \citep{Schodel+2020} it would not be implausible to observe a 3BPU merger product in the more central regions of the NSC. Directly observing a 3BPU would be unlikely; less so than observing a normal merger or collision, which is already extremely rare. Also, when so close to the SMBH, ZLK-induced collisions may deplete the available sample of binaries too quickly for slower scatter-induced 3BPUs to occur.  

The middle panel of Figure \ref{fig:3BPUs} shows the difference between the actual age $t_{\rm f}$ (e.g. the time of the 3BPU) and the apparent age $\tau_{\rm age}$ of the 3BPU product, using the helium core mass fraction to infer the age. This assumes there is no mixing, and that the new age is always determined by the most massive star. The difference should be an estimate of how much younger (or older) the 3BPU product would appear if observed. Although we estimate that most of the 3BPUs would appear a few $\mathrm{Gyr}$ younger, we also estimate differences almost as high as $10~\mathrm{Gyr}$. Such a discrepancy could suggest observations of very young and metal-poor stars in the NSC, similar to, if not even more extreme, than the BSS candidate in \citet{Thorsbro+2023}.

Finally, the bottom panel of Figure \ref{fig:3BPUs} shows the distribution of the final masses; apart from the more massive stars $> 2~\mathrm{M}_\odot$ the 3BPUs populate the region just below $ 2~\mathrm{M}_\odot$ as well, which in part explains the peak seen around this value in Figure \ref{fig: merger orbits hists}. The details of the 3BPU merger products from these simulations are rather limited. We assume that the initial collision conserves mass and angular momentum, which is not a bad approximation since we do not expect to have any collisionally stripped stars \citep[see][and Figure \ref{b distribution}]{Rose+2023}, but means that the mass distribution is an upper limit. Second, grazing collisions, where the periapsis approach is similar to the sum of the stellar radii ($r_p \sim R_{ij}$) do not necessarily result in the coalescence of the two stars; there would simply be no 3BPU, and as such the number/fraction of 3BPUs is also an upper estimate. We also made the assumption of no mixing for stellar collisions and mergers. The amount of mixing that occurs as a result of the coalescence of two stars depends on various mechanisms: the relative velocity of the colliding stars, the impact parameter, the stars' initial masses and chemical compositions (hence also evolutionary stage), the mass loss of the collision etc. \citep[see e.g.][]{Sills+1997,Gibson+2025}. Head-on collisions between stars in globular clusters, with masses similar to those considered in our simulations, have been shown not to induce additional convection, thereby limiting the amount of fresh burning material transported into the core region \citep{Sills+1997, Sills_2009}. \citet{Sills+2001} showed that BSSs that rotate typically remain longer on the MS than their non-rotating counterparts. Off-axis collisions can increase the rotational speed of the collision product, which in turn leads to rotation-induced mixing.

The masses and stellar lifetimes of our collision and merger products are undoubtedly affected by these mechanisms, but should not affect our results in the broader context. The middle and bottom panels of Figure \ref{fig:3BPUs} are only rough estimates. The precision of our results could be enhanced by making hydrodynamic simulations of the stellar mergers and incorporating stellar evolution.

\REV{
\subsection{Comparisons to other studies} \label{Discussion: Other studies}

We have previously mentioned a few studies that also looked at binaries and their evolution in the NSC; \citet{Rose} conducted an analytical where the shortest timescale determines the fate of a binary, and \citet{Panamarev+2019} carried out the largest $N$-body simulation of the NSC to date. Below, we limit the discussion to these two studies as they are the most relevant to ours, highlighting their most relevant results and how they compare with ours.

Our findings are overall in good agreement with \citet{Rose}; initially softer binaries evaporate, and those that are not, typically remain stable against evaporation. We find that the rate of evaporation is somewhere in-between their lower and upper limits (see e.g. the evaporation and max-evaporation curves in their figure 1), with a typical evaporation timescale of a few $\sim 100~\mathrm{Myr}$. Binaries that do not evaporate, as in those that initially are marginally soft or hard, will most likely merge (or alternatively migrate inward where they can evaporate) on a timescale set by the ZLK quadrupole timescale or the segregation timescale, which agrees with our results, see Figure \ref{fig: ZLK v Scat times}. \cite{Rose} suggest that migration plays a secondary role to evaporation in shaping the binary distribution of the NSC. Although we also find that evaporation dominates (along with mergers), we would like to put extra emphasis on the migration's role in shaping the binary distribution in the NSC. For example, the widest possible separation a $2~\mathrm{M}_\odot$ binary could have after $1~\mathrm{Gyr}$ at $0.3~\mathrm{pc}$ is estimated to be $\sim 0.1~\mathrm{au}$ \citep[][Fig. 7]{Rose} and is set by the evaporation timescale. We, however, find a handful of such binaries at the same distance even after $10 ~\mathrm{Gyr}$, where some originate from $0.1~\mathrm{pc}$. 

Furthermore, binaries may oscillate stochastically in their distance to the SMBH, which gives them a chance to momentarily harden (soften) further out (in) in the NSC \citep[consider again Heggie's law,][]{Heggie1975}. When it comes to the limits, e.g. the maximum separation we can expect at a given distance to the SMBH \citep[which essentially is the purpose of][]{Rose}, migration, or these stochastic oscillations, become very important because it significantly changes the correlation between inner and outer orbits as a function of time. Finding a wider binary than expected does therefore not imply a recent dynamical formation scenario for the binary, nor does it exclude it. 

\citet{Panamarev+2019} ran an $N$-body simulation of the NSC for $\sim 5~\mathrm{Gyr}$ with $10^6$ particles and an initial binary fraction of 5\%. Their initial binaries differ from ours in a few ways; they consider stellar masses of $0.08-100~\mathrm{M}_\odot$, binary separations between $0.005-50~\mathrm{au}$ and more generally they cover the inner $10~\mathrm{pc}$ of the NSC. Qualitative trends are the same; wide binaries become less common with time as they are either destroyed or become harder, the binary fraction has a radial dependence (e.g. more binaries remain intact further out), and binaries $\gtrsim1~\mathrm{au}$ become very rare. In particular, they find that initially wide (soft) binaries, apart from evaporating, also become harder (see their Fig. 10). What is surprising, in comparison to our results, is the large number of hard binaries they obtain. In absolute terms, there is roughly a factor of 3 more hard binaries than in their initial population, with a maximum around $a_\mathrm{bin}\sim2 ~\mathrm{R}_\odot$ (dominated by low mass MS binaries). Overall, they find that about half of the initial binary population is destroyed after $\sim 5~\mathrm{Gyr}$ (30-40\% are therefore hardened). Although this is dependent on their location in the NSC, this seems to be the case in the inner 1--0.1 parsec (see their fig. 9 and 12). In contrast, we find a significantly stronger depletion of binaries, along with a more pronounced dependence on distance from the SMBH over time, with roughly 80\% (60\%) originating from $0.1~\mathrm{pc}$ ($0.3~\mathrm{pc}$) being destroyed after $\sim 5~\mathrm{Gyr}$, see e.g. Fig. \ref{fig: evolution1}.
There are, however, a few possible explanations for these discrepancies: \textit{i)} The binary evolution code \texttt{BSE} \citep{Hurley+2002} used in their code has very efficient tidal circularisation for fully convective stars. \textit{ii)} Their paper lacks sufficient detail to determine precisely how binaries are treated, but it appears that the overall encounter rate is likely underestimated due to each N-body particle representing 65 stars. This would lead to an underestimation of binary destruction (e.g. through evaporation) and therefore alter the properties of the surviving binary population. \textit{iii)} other relaxation processes not considered in our simulations could change the statistics of our outcomes, see a discussion in Section \ref{Section: VRR}.

\subsection{Resonant relaxation} \label{Section: VRR}

Certain dynamical processes that were not considered in our simulations could have an affect on our results. Non-spherical fluctuations in the overall stellar distribution lead to torques acting on the stellar orbits; these torques drive diffusion in both the direction and the magnitude of the orbital angular momentum of the stellar orbits around the SMBH \citep{KocsisANDTremaine2015, Hamers+18, Tep+21, PanamarevANDKocsis2022}. The random re-orientation of the direction of the angular momentum is known as vector resonant relaxation (VRR), while changes to the magnitude occur on longer timescales and is known as scalar resonant relaxation (SRR). The main effect of SRR is that the overall eccentricity distribution of the binaries' outer orbits is driven to be thermal; as we already assume a thermal distribution, this would simply add an ever randomising effect on their outer eccentricities. This is most likely not going to affect the statistics of our results much. ZLK is operating on shorter timescales for virtually all distances considered (including migrated binaries) and there is only a weak dependence on outer orbit eccentricities when it comes to evaporation \citep{Rose}.

The effects of VRR are likely more relevant; a constant re-orientation of the outer inclination (e.g. direction of angular momentum) has been shown to increase the number of mergers via the ZLK mechanism \citep[the mutual inclination of a binary can sporadically reach $\sim 90$ degrees where the ZLK oscillations are the strongest][]{Hamers+18}. At the same time, \citet{Dodici+2025} show that the combination of VRR, ZLK, flybys via the impulse approximation and strong tides can be very effective in circularising binaries, making them virtually stable against any continued ZLK as well as evaporation. It is important to note that this primarily affects our initially softer binaries since the harder binaries do not undergo ZLK to begin with. In other words, the binaries initially near the H/SB are unlikely to be affected significantly. How important these processes are therefore depends on the initial distribution one considers. Overall, our conclusion is that VRR likely increases the mergers and the amount of binaries that remain intact, while the number of evaporating binaries decreases. A comprehensive study of the complex interplay is left for a later investigation.}

\section{Conclusions}\label{Conclusion}

In this work, we have followed the dynamical evolution of 5736 lower-mass ($\le 2~\mathrm{M}_\odot$) binaries initially at $0.1 ~\& ~0.3~\mathrm{pc}$ from the SMBH in the NSC. The dynamical evolution entails close encounters with field stars in the NSC, ZLK oscillations induced by the SMBH, tidal dissipation, and stellar evolution by interpolating pre-computed tracks. We find that binary migration caused by three-body encounters plays a vital role in determining the outcome of binaries; inward migration leads to a certain death in the form of mergers and evaporation, while outward migration serves as a way to retain binaries within the NSC. We pay extra attention to the stellar collisions and binary mergers in these simulations because of their likely connection to the G-objects in the inner parts of the NSC. Although we find that binaries, as well as their merger products, can end up on orbits confined to this region -- which suggests that the S-cluster and the G-objects can originate further away from the SMBH -- the time it takes for the binaries to end up there is too long ($0.3-1.2~\mathrm{Gyr}$) compared to the apparent ages of these objects (${\sim} 10~\mathrm{Myr}$). \REV{Instead, we suggest that the S-cluster may be contaminated by low-mass BSS that are related to earlier SFEs in the NSC \citep{Schodel+2020}.}

We also find that in a majority ($\lesssim80\%$) of cases where a surrounding star collides with one of the binary stars, it leads to a subsequent merger with the remaining binary star, resulting in a three-body pile up (3BPU). These are relatively common within the first 1-2 Gyr but stagnate later on. Based on a simple interpolation of the core-helium mass fraction of the 3BPU products, these objects could appear up to $10~\mathrm{Gyr}$ younger than their true age. 
Finally, with a SFE possibly having occurred ${\sim}2~\mathrm{Gyr}$ ago \citep{Lu+2013,Schodel+2020}, we would expect to see both recent mergers (e.g. in the form of G-like objects) and more evolved merger products \citep[e.g. in the form of][-like BSS candidates]{Thorsbro+2023} associated to this SFE, scattered throughout the NSC as a consequence of binary migration.

\begin{acknowledgements}

\REV{We would like to thank the anonymous referee for their comments and suggestions that greatly improved this paper.} We give special thanks to Abbas Askar and Alessandra Mastrobuono-Battisti for their comments and valuable suggestions, which also helped improve this paper.
The computations and data handling were enabled by resources provided by LUNARC, The Centre for Scientific and Technical Computing at Lund University. AAT acknowledges support from the Horizon Europe research and innovation programs under the Marie Sk\l{}odowska-Curie grant agreement no. 101103134.
This work made use of Astropy: a community-developed core Python package and an ecosystem of tools and resources for astronomy \citep[][\url{http://www.astropy.org}]{astropy:2013, astropy:2018, astropy:2022}; Scipy: \citep[][\url{https://scipy.org}]{2020SciPy-NMeth}, Numpy: \citep[][\url{https://numpy.org/}]{harris2020array}; Matplotlib: \citep[][\url{https://matplotlib.org}]{Hunter:2007} and Pandas: \citep[][\url{https://pandas.pydata.org}]{mckinney-proc-scipy-2010}

\end{acknowledgements}

\section*{Data Availability}
The data from this article will be shared upon reasonable request.

\bibliographystyle{aa}
\bibliography{references}

\begin{appendix}

\section{Accept/Reject method} \label{Accept/Reject}
\begin{figure*}[htbp]
    \centering
    \includegraphics[width=0.9\linewidth]{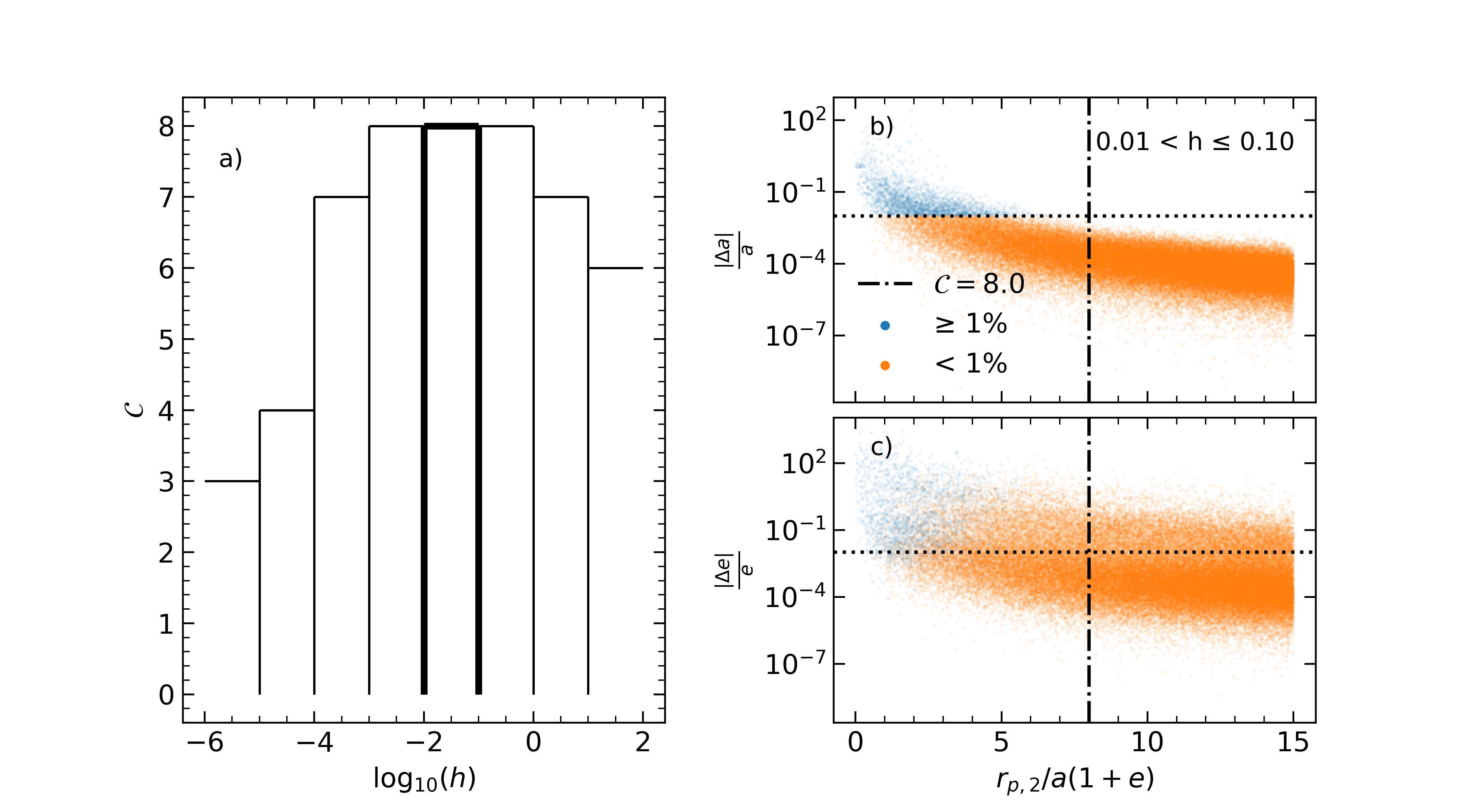}
    \caption{{\it Left:} Scaling parameter as a function of hardness. The edges are open to illustrate that any binary with, for example $h > 10$ has $\mathcal{C} = 6$. {\it Right:} Scatter plots for the fractional change in separation of binaries (top, panel b) and eccentricity (bottom, panel c), against the distance between the tertiary and binary centre-of-mass in units of the binary apoapsis. Only encounters with with a hardness of $0.01 < h \leq 0.10$ are considered in this Figure. Blue (orange) dots correspond to encounters that change the binary separation with $\geq 1\%$ ($< 1\%$), where the black dotted lines show the $1\%$ boundary and the black dash-dotted line the chosen value for the scaling parameter $\mathcal{C}$.}
    \label{fig: Scaling Parameter}
\end{figure*}
\citet{Alessandra} constructed a method to sample encounters that lead to collisions between two single stars. We modify this method to handle interactions between a binary system and a single star. Binaries predominantly evolve through scatters, so the cross-sections considered for these three-body encounters are not confined to the physical sizes of the stars themselves. We first define the effective two-body periapsis distance $r_{\rm p,2}$ as the closest approach between the tertiary and the centre of mass of the binary, assuming that the latter is a point mass. We then consider encounters out to a maximum distance of $r_{p,2}^\mathrm{max} = \mathcal{C}a(1+e)$, where $a$ is the binary separation, $e$ is the eccentricity, and $\mathcal{C}$ is a scaling parameter. Generally, the farther away the periapsis approach of the tertiary is from the binary, the less the binary will be affected. The scaling parameter helps reduce encounters that have no effect on the binary. We define $\mathcal{C}$ empirically as a function of a binary's hardness; out to how many apoapsis do we expect encounters to induce changes $> 1\%$ to the binary's separation. Therefore, it may be thought of as the parameter that maximises the effective two-body periapsis distance ($r_{p,2}$) between the tertiary and the binary while keeping the number of interactions at a minimum. The empirical values and details on how they are obtained can be found in \ref{Empirical Study}.

For an encounter with total binary mass $M_{1,2}$, tertiary mass $M_3$, relative velocity at infinity $v_\infty$, and impact parameter $b$, one can show that

\begin{equation} \label{periapsis for C}
    r_{p,2} = -\frac{G(M_{1,2}+M_3)}{2v_\infty^2}\left[1-\sqrt{1 + \left(\frac{2v_\infty^2}{G(M_{1,2}+M_3)}\right)^2b^2}\right],
\end{equation}
where $G$ is the gravitational constant.

With empirically determined values for $\mathcal{C}$, we construct an accept-reject method for binary-tertiary encounters based on a modification of the approach in \citet{Alessandra}. First we sample a “velocity at infinity” from a Maxwellian distribution with a velocity dispersion according to eqn. (\ref{Eq: Velocity Dispersion}), which at $0.1~\mathrm{pc}$ gives $\sigma \sim 250~\mathrm{kms}^{-1}$. We then define the encounter rate $\mathcal{R}_{i j}^v$ \citep[see eqn. \ref{Eq: Encounter Rate} and][eqn. 13]{Alessandra} where $n_j$ is the number density, $M_i$ is the binary mass, $M_j$ is the tertiary mass and $R_{i j} = \mathcal{C} a(1+e)$ is the maximised effective two-body periapsis (e.g. $r_{p,2}^\mathrm{max}$). We may choose a time-step $\Delta t$ such that there is an interaction when $X < \mathcal{R}_{i j}^v \Delta t$, where $X$ is a randomly chosen number between 0 and 1. The associated “average” timescale for an encounter is given by eqn. (\ref{eqn: tenc}); where we have replaced $v_\infty$ with the velocity dispersion $\sigma(r_\bullet)$.
\begin{equation} \label{Eq: Encounter Rate}
\mathcal{R}_{i j}^v\left(v_{\infty}\right)=n_j \pi R_{i j}^2 v_{\infty}\left[1+\frac{2 G\left(M_i+M_j\right)}{v_{\infty}^2 R_{i j}}\right]
\end{equation}
Once we accept an interaction we can calculate the interaction radius, see eqn. (\ref{collision radius}), where $R^2_\mathrm{rand} = X/(\pi n_j v_\infty \Delta t)$ and $R_\mathrm{esc} = 2G(M_i + M_j)/v_\infty^2$. The interaction radius corresponds to the collision radius in \citet[eqn. 15 ][]{Alessandra}.

\begin{equation} \label{eqn: tenc}
    \bar{t}_\mathrm{enc} = \frac{1}{2\pi}\left(4 \sqrt{\pi}\sigma n_j R_{ij}^2 (1 + \frac{M_{ij}}{2\sigma^2 R_{ij}}) \right)^{-1}
\end{equation}

Using $R_\mathrm{col}$ and conservation of energy and angular momentum, we can infer the initial impact parameter of the tertiary using eqn. (\ref{b distribution}). 
\begin{equation} \label{collision radius}
    R_\mathrm{col} = \frac{R_\mathrm{esc}}{2} \left[ \sqrt{1 + 4\left(\frac{R_\mathrm{rand}}{R_\mathrm{esc}}\right)^2} -1\right]
\end{equation}
\begin{equation}\label{b distribution}
    b^2 = R_\mathrm{col}^2 \left[ 1 + \frac{2G(M_i+M_j)}{v_\infty^2 R_\mathrm{col}} \right].
\end{equation}

\subsection{Setting up a tertiary} \label{Setting up a tertiary}

After having sampled a velocity at infinity $v_\infty$ and an impact parameter $b$ (see Appendix \ref{Accept/Reject}) for the tertiary, we can determine the position and velocity of the tertiary. Although $v_\infty$ and the impact parameter are sampled at infinity, these quantities need to be extrapolated to a finite distance from the binary; the binary should also be in a random position in its orbit at the time of closest approach. As such, we placed the tertiary at a distance $D$ where, at a constant velocity $v_\infty$, it takes 10 binary orbital periods to reach the binary. The extrapolation is achieved using conservation of energy and angular momentum in which we treat the binary center of mass as a point mass. Denoting the extrapolated velocity and impact parameter by $v_\mathrm{D}$ and $b_\mathrm{D}$ respectively, we get
\begin{equation} \label{Eq: v effective}
    v_\mathrm{D} = \sqrt{v_\infty^2 + \frac{2GM_{ij}}{\sqrt{D^2 + b_\mathrm{D}^2}}}; \quad b_\mathrm{D} = \frac{bv_\infty}{v_\mathrm{D}}
\end{equation}
With $v_\mathrm{D}$ and $b_\mathrm{D}$, we can construct the tertiary's initial 6D phase-space coordinates (position and velocity vectors). We begin by drawing a value for an angle $\phi \in [0, 2\pi]$ and $z \in [-1,1]$, both from uniform distributions. We then define $R = \sqrt{1 - z^2}$ from which it follows that $x = R \sin{\phi}$ and $y = R \cos{\phi}$. The initial position $\mathbf{r}$ without any offset is simply given by the three coordinates $x,y,z$ multiplied by the distance $D$. To set up the offset vector $\mathbf{b}$ we first make a coordinate transformation and define $\mathbf{\hat{i}} = \frac{\mathbf{\hat{x}}\cross \mathbf{\hat{r}}}{|\mathbf{\hat{x}}\cross \mathbf{\hat{r}}|}$ \footnote{if $\mathbf{\hat{r}} \parallel \mathbf{\hat{x}}$ we make the change $\mathbf{\hat{x}} \rightarrow \mathbf{\hat{y}}$.}. We then get the impact parameter vector
\begin{equation}
    \mathbf{b} = \left(\mathbf{\hat{i}}\sin{\phi_2} + \mathbf{\hat{j}}\cos{\phi_2}\right) b_\mathrm{D}
\end{equation}
where $\mathbf{\hat{j}} = \mathbf{\hat{i}} \cross \mathbf{\hat{r}}$ and $\phi_2$ is drawn from a uniform distribution $\phi_2 \in [0,2\pi)$. The phase-space coordinates with the offset are given by
\begin{equation} \label{Eq: Phase Space coordinates for star 3}
\left\{\begin{array}{cc}
 \mathbf{x} =& D\mathbf{r} + \mathbf{b}  \\
 \mathbf{v} =& -v_\mathrm{D}\mathbf{\hat{r}} \\
\end{array}\right.
\end{equation}

\noindent Lastly, the tertiary's mass is drawn from a two-part power law initial mass function (IMF) with $\alpha = -1.3$ for masses below $0.5~\mathrm{M}_\odot$ and $\alpha = -2.3$ for masses above \citep{Kroupka2001}. We restrict the tertiary mass to $0.1 < m_3 /\mathrm{M}_\odot < 1.0$, where the upper constraint is imposed from the turn-off mass ($\sim 1~\mathrm{M}_\odot$) and all stars considered are on the MS throughout the simulation. For simplicity we adopt a fixed upper limit for $m_3$ which causes us to neglect a smaller number of encounters in the early life of the binary in which the tertiary could be more massive than the primary.

\subsection{The scaling parameter $\mathcal{C}$} \label{Empirical Study}

In order to reduce the computational cost for these simulations, we have introduced a scaling parameter $\mathcal{C}$ that maximises the effective two-body periapsis distance ($r_{p,2}$) between the tertiary and the binary while keeping the number of interactions at a minimum. By matching the inferred pericentric passage in eqn. (\ref{periapsis for C}) to the induced changes in a binary's semi-major axis and eccentricity we can empirically determine values of $\mathcal{C}$ for different hardness ratios. In total, we ran $10^6$ individual interactions for this empirical test; each encounter samples a tertiary and a binary in the same way as for the real simulations, see Section \ref{Stellar Interactions}. 

\begin{figure}[H]
    \centering
    \includegraphics[width=0.95\linewidth]{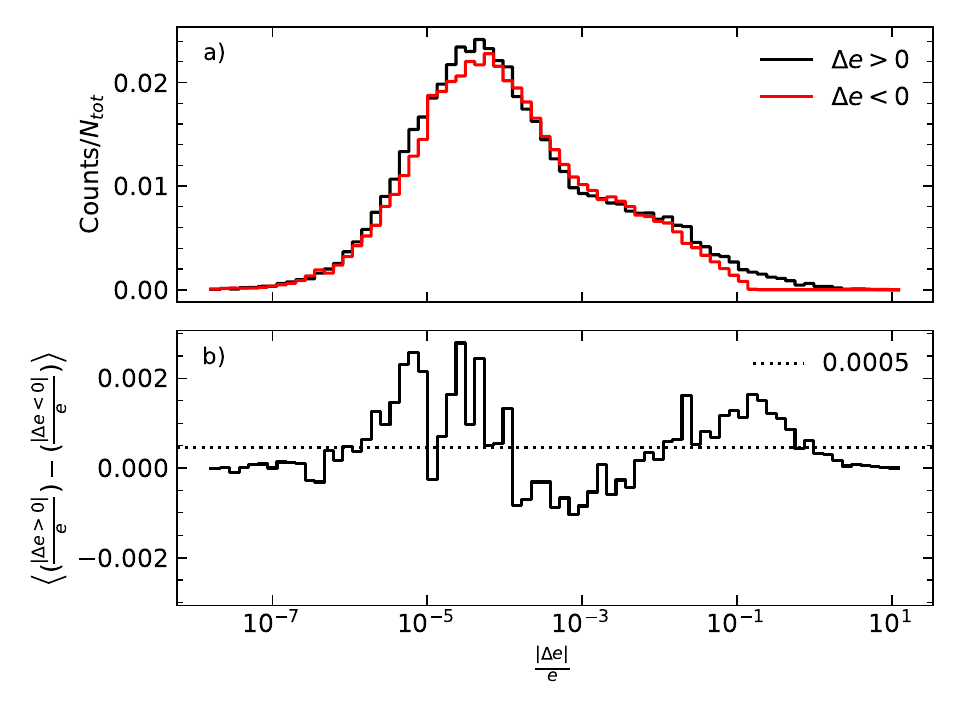}
    \caption{{\it Top:} Distribution of the fractional effect on a binary's eccentricity for encounters that increase (black line) and decrease (red line) it. {\it Bottom:} Residuals for the top panel. The dotted black line shows the average (0.0005) from the residuals.}
    \label{fig: eccentricity changes}
\end{figure}

The left-hand side panel in Figure \ref{fig: Scaling Parameter} shows our chosen values as a function of the logarithm of the hardness ratio of the binary. The values of $\mathcal{C}$ for the smallest and biggest hardness ratios are not closed (e.g. $h=10^2$ gives the same scaling parameter as $h=10^4$). The number of encounters that fall in these brackets is virtually zero and is expected to have no effects on the main simulations. One of the brackets ($0.01 < h \leq 0.10$) is also highlighted; the two panels on the right show the outcomes of each encounter for these hardness ratios. In the top (bottom) panel, we plot $|\Delta a/a|$ ($|\Delta e/e|$) against $r_{p,2} / a(1+e) \equiv \mathcal{C}$. Distant encounters are more likely to cause significant changes in a binary's eccentricity than in its separation. This is mainly due to a larger spread, making it more difficult to justify specific values for $\mathcal{C}$. For example, at $\mathcal{C}=8$ (chosen value for the bracket $0.01 < h \leq 0.10$), there are no encounters that induce changes $0.01 < |\Delta a|/a$, but there are many encounters that induce changes $0.01 < |\Delta e|/e$. We make the assumption that it is enough to consider the changes in separation for the scaling parameter. This means that we are neglecting multiple strong encounters with respect to changes in eccentricity. However, as we have roughly equally many encounters that decrease the eccentricity as there are encounters that increase it, we argue it is a reasonable assumption to make. We show this in Figure \ref{fig: eccentricity changes}; in the top panel we have the distributions of encounters ($>\mathcal{C} = 8$ for $0.01 < h \leq 0.10$) that increase the eccentricity (black curve) and encounters that decrease it (red curve), normalised to the total number of encounters. In the bottom panel, we show the difference between the two distributions. On average, there is an increase of $0.0005$ associated to each encounter. We are therefore underestimating the rate at which the eccentricity increases, but only by a very small amount.

\section{Secular processes} \label{Appendix: Secular Processes}

The Hamiltonian of a hierarchical triple can be decomposed into the inner orbit (e.g. the binary), the outer orbit (e.g. the binary centre of mass around the SMBH) and a coupling term describing the weak interaction between the orbits. The corresponding Hamiltonian is given by eqn. (1) in \citet{Naoz+2013a}, and is defined to be the negative of the total energy, following the convention of \citet{Harrington1968}. \citet{Naoz+2013a} adopt the canonical variables known as Delaunay's elements; these coordinates are chosen to be the mean anomalies $l_1$ and $l_2$, the longitudes of ascending nodes $h_1$ and $h_2$ and the argument of periapsis $g_1$ and $g_2$. Subscripts 1 and 2 denote the inner and outer orbits, respectively. For the full set of differential equations, we refer the reader to \citet[][see their Section 3 and Appendix A]{Naoz+2013a}. We choose to integrate the differential equations of the Keplerian orbital elements \citep[][see Eqns. A29-A35]{Naoz+2013a}. We explicitly state the expression for the evolution of the semi-major axis in eqn. (\ref{Eq: dadt ZLK}) which is not given in the original paper. This expression can be obtained by combining the conjugate momenta of the mean anomalies ($L_1$) and the arguments of periapsis ($G_1$), see eqns. (3) and (4) in \citet{Naoz+2013a}, respectively.

\begin{equation} \label{Eq: dadt ZLK}
    \Dot{a}_1 = \frac{\sqrt{(m_1+m_2)a_1} \left( \Dot{G}_1 + L_1\frac{e_1}{\sqrt{1-e_1^2}}\Dot{e}_1\right)}{m_1m_2\sqrt{1-e_1^2}}.
\end{equation}

Here $m_1,~m_2$ are the stellar masses of the binary stars, and $a_1, ~e_1$ are the separation and eccentricity of the inner orbit, respectively.

\section{Power-law fits}
\begin{table}[H]
    \caption{Power-law fits for the eccentricity distribution of the different outcomes.}
    \label{table: power law fits} 
    \centering
    \begin{tabular}{c c c c c} 
        \hline\hline
        
        Outcome & $\eta_{in, 0.1}$ & $\eta_{in, 0.3}$ & $\eta_{out, 0.1}$ & $\eta_{out, 0.3}$ \\
        \hline
        
        Intact & -0.4447 & -0.4430 &  1.7363 &  1.6728 \\
        Evolved & -0.4979 & -0.3990 &  1.1165 &  0.9696 \\
        Evaporated &  0.2399 &  0.3745 &  0.8965 &  1.1062 \\
        Merged &  8.0233 &  5.6852 &  1.3190 &  1.4793\\
        Ejected & -0.3532 & -0.4359 &  0.9650 &  1.5411\\ \hline
        Initial & -0.4562 & -0.4561 &  1.0000 &  1.0000
        
    \end{tabular}

\end{table}

\section{Tidal disruption events} \label{Appendix TDE}

\begin{figure}[H]
    \centering
    \includegraphics[width=0.9\linewidth]{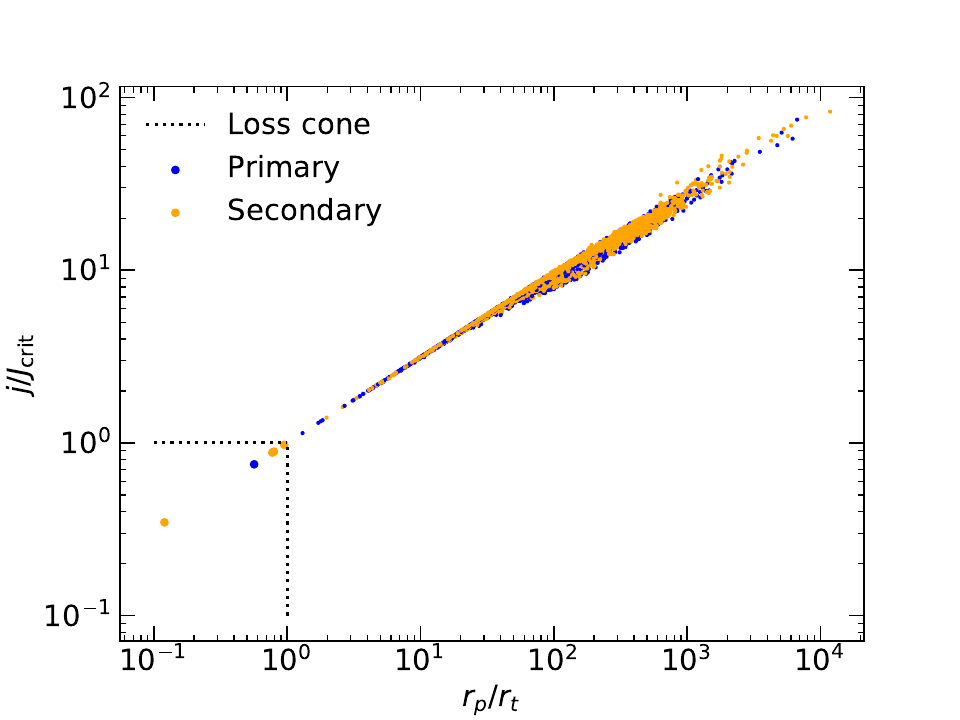}
    \caption{Ratio between the periapsis and tidal radius versus the ratio between the specific angular momentum and the critical specific angular momentum for the loss-cone. Blue points correspond to the primary of the binary while orange points correspond to the secondary. The dotted black lines shows where respective ratio is unity - values within the box are inside the loss-cone.}
    \label{fig: disrupted stars}
\end{figure}

When binaries evaporate, the now single stars typically find themselves on eccentric bound orbits around the SMBH. There is then the possibility that they venture close enough to the SMBH to be tidally disrupted. A tidal disruption event (TDE) requires the periapsis to be smaller than the star's tidal radius, and its angular momentum to be smaller than the critical angular momentum associated with the loss cone. Figure \ref{fig: disrupted stars} shows the ratio between the periapsis and the tidal radius along the x-axis and the ratio between the star's angular momentum and the critical angular momentum along the y-axis. We find a total of five possible TDEs from these simulations.

\begin{figure}[H]
    \centering
    \includegraphics[width=\linewidth, trim={0 2cm 0 3cm},clip]{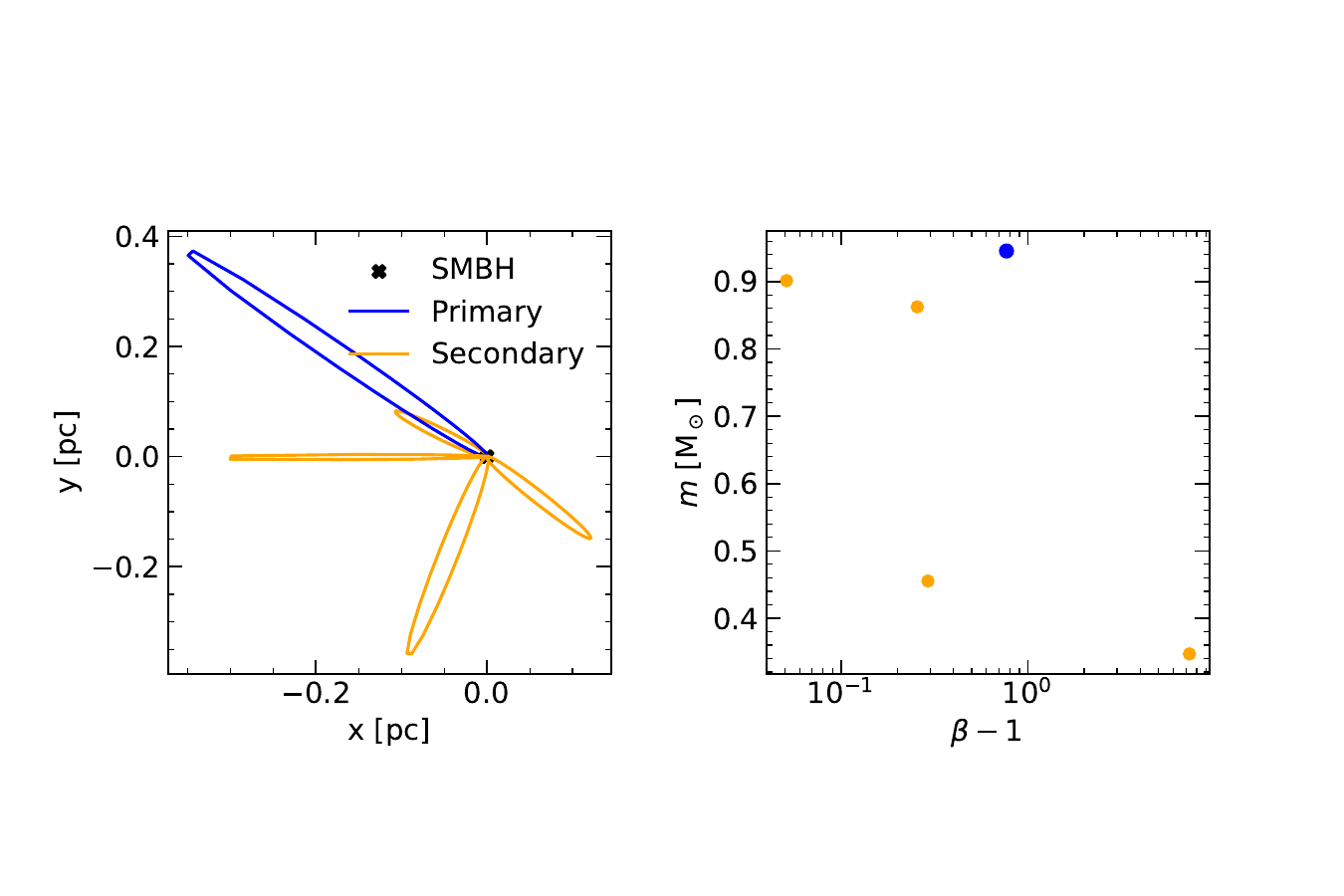}
    \caption{{\it Left:} Orbits for the stars found inside the loss cone in Figure \ref{fig: disrupted stars} with the SMBH at the origin. {\it Right:} Scatter between the stars' mass and the penetration parameter $\beta \equiv \frac{r_t}{r_p}$. The orbits and points share the same colour-coding as Figure \ref{fig: disrupted stars}.}
    \label{fig: loss cone}
\end{figure}

Figure \ref{fig: loss cone} shows the stellar orbits around the SMBH (origin) in the left-hand side panel. The right-hand side panel shows the penetration parameter of a TDE $\beta \equiv \frac{r_t}{r_p}$, where $\beta - 1 \lesssim 1$ corresponds to non-violent events that may lead to partial disruption of the stars \citep[e.g.][]{Phinney1986, Guillochon+2013}. While the number of TDEs is far too few to infer any statistical properties of these events, we can say that the disruption of binaries are going to lead to TDEs, and that they are most likely going to be non-violent. 

\section{Inner arcsec orbits} \label{Appendix: Inner arcsec orbits} 

\begin{figure}[H]
    \centering
    \includegraphics[width=0.95\linewidth]{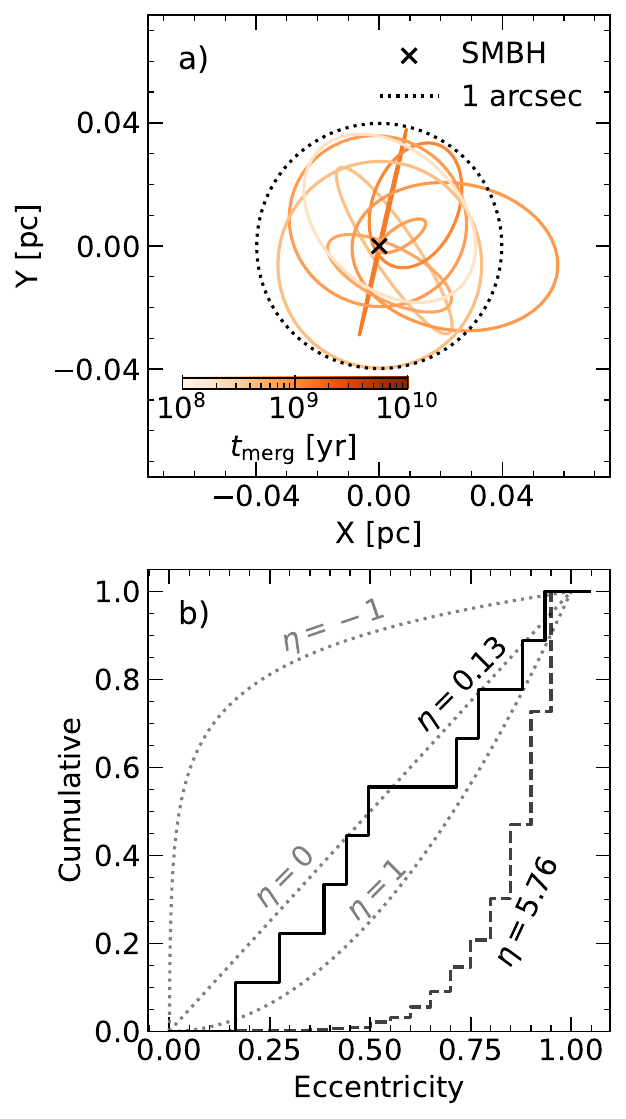}
    \caption{{\it Top:} 2D projection of the 9 merger products' orbits colour-coded as a function of time of merger. The black cross in the origin corresponds to the SMBH and the dotted line marks the inner arcsec. {\it Bottom:} Cumulative distribution of the merger products' eccentricities around the SMBH, similar to Figure \ref{fig: ecc distr}. The solid black line corresponds to the merger orbits seen in the top panel, while the dashed black line corresponds to all mergers with a periapsis $a(1-e) \le 0.04~\mathrm{pc}$.}
    \label{fig:inner arcsec}
\end{figure}

Figure \ref{fig:inner arcsec} shows a 2D projection of all merger products' (see Figure \ref{fig: merger orbits hists}) orbits with a semi-major axis smaller than $0.04~\mathrm{pc}$ which corresponds to the inner arcsec of the GC (colour shows the time at which they merge); this is also more or less the region where the S-stars reside. We find that 9 binaries merge with orbits in this region, and migrate there within a few $100~\mathrm{Myr}$ to a bit more than $1~\mathrm{Gyr}$ (mean time is $660~\mathrm{Myr}$). All of the mergers within the inner arcsec originate from $0.1~\mathrm{pc}$ which roughly corresponds to $1\%$ of the mergers from this distance. In total there are 14 and 3 binaries originating at $0.1~\mathrm{pc}$ and $0.3~\mathrm{pc}$, respectively, that at some point in their lives have an orbit with $a_\mathrm{outer}<0.04~\mathrm{pc}$; those not included in the 9 mergers above, merge at a slightly greater distance. None of the additional binaries migrate into the region quicker than $300~\mathrm{Myr}$.
The bottom panel of the figure shows the cumulative distribution of eccentricities. The best power-law fit (solid black line) is $\eta \approx 0.08$, which is surprisingly uniform in contrast to the eccentric orbits of the S-stars \citep{2009ApJ...707L.114G, Gillessen+2009, Sabha+2012} and the G-objects \citep{Ciurlo+2020}. In the same panel, the dashed black line shows a super-thermal distribution for all binary merger orbits with a periapsis within the inner arcsec, which roughly corresponds to 60\% of the mergers originating at $0.1~\mathrm{pc}$. The uniform distribution may therefore, or at least partially, be caused by a low number of orbits. 

\end{appendix}

\end{document}